\documentclass[12pt]{article}

\usepackage{amsbsy}
\usepackage{amsfonts}
\usepackage{amssymb}

\usepackage{bm}
\usepackage{enumerate}
\usepackage{graphicx}   % choix graphicx au lieu de graphics, car figures en format .png (plus petit que .eps)
\usepackage{epsfig}
\usepackage{color}
\usepackage{nicefrac}
\usepackage{mathrsfs}
\usepackage{amsmath}
\usepackage{amsthm}
\DeclareFontFamily{OT1}{pzc}{}
\DeclareFontShape{OT1}{pzc}{m}{it}{<-> s * [1.200] pzcmi7t}{}
\DeclareMathAlphabet{\mathpzc}{OT1}{pzc}{m}{it}
\input epsf.sty

\newlength{\TZ}
\newcommand{\BEQ}{\begin{equation}}     % Gleichungen Anfang ..
\newcommand{\BEA}{\begin{eqnarray}}
\newcommand{\BD}{\begin{displaymath}}
\newcommand{\EEQ}{\end{equation}}       % .. und Ende
\newcommand{\EEA}{\end{eqnarray}}
\newcommand{\ED}{\end{displaymath}}
\newcommand{\eps}{\varepsilon}          % epsilon
\newcommand{\vep}{\varepsilon}          % epsilon
\newcommand{\vph}{\varphi}              % rundes phi
\newcommand{\D}{{\rm d}}                % gerades d fuer Ableitungen
\newcommand{\II}{{\rm i}}               % gerades i fuer komplexe Einheit
\renewcommand{\lg}{{\rm lg\,}}          % dekadischer Logarithmus (instruction \lg renouvellee !!)
\newcommand{\demi}{\frac{1}{2}}         % Bruch 1/2
\newcommand{\wit}[1]{\widetilde{#1}}    % weite Schlange
\newcommand{\lap}[1]{\overline{#1}}     % Querstrich oben
\renewcommand{\vec}[1]{\boldsymbol{#1}} % Vektoren fettgedruckt

\newcommand{\fnref}[1]{$^{\ref{#1}}$}   % marquer pour une note de bas de page (footnote) 
\catcode`\@=11
\def\numberbysection{\@addtoreset{equation}{section}
        \def\theequation{\thesection.\arabic{equation}}}
\numberbysection

\definecolor{gruen}{rgb}{0,0.625,0}     % semble une solution pour obtenir un vert assez sombre
\definecolor{rot}{rgb}{0.75,0,0}        % essai pour obtenir un rouge sombre
\definecolor{blau}{rgb}{0,0,0.75}       % essai pour obtenir un bleu sombre
\newcommand{\ROT}[1]{\textcolor{rot}{{\rm #1}}}	    % Texte en rouge sombre
\newcommand{\VERT}[1]{\textcolor{gruen}{{\rm #1}}}	% Texte en vert sombre
\newcommand{\BLAU}[1]{\textcolor{black}{{\rm #1}}}	% Texte remis en noir pour version a publier

\begin{document}

\begin{titlepage}

\vskip 1.5 cm
\begin{center}
{\LARGE \bf Infinite-dimensional meta-conformal Lie algebras in one and two spatial
dimensions\footnote{{\it In memoriam} Vladimir Rittenberg}}
\end{center}

\vskip 2.0 cm
\centerline{{\bf Malte Henkel}$^{a,b}$ and {\bf Stoimen Stoimenov}$^c$}
\vskip 0.5 cm
\centerline{$^a$ Laboratoire de Physique et Chimie Th\'eoriques (CNRS UMR 7019), Universit\'e de Lorraine Nancy,}
\centerline{B.P. 70239, F -- 54506 Vand{\oe}uvre l\`es Nancy Cedex, France}
\vspace{0.5cm}
\centerline{$^b$ Centro de F\'{i}sica Te\'{o}rica e Computacional, Universidade de Lisboa, P -- 1749-016 Lisboa, Portugal}
\vspace{0.5cm}
\centerline{$^c$ Institute of Nuclear Research and Nuclear Energy, Bulgarian Academy of Sciences,}
\centerline{72 Tsarigradsko chaussee, Blvd., BG -- 1784 Sofia, Bulgaria}

\begin{abstract}
\BLAU{Meta-conformal transformations are constructed as sets of time-space transformations which are not angle-preserving but 
contain time- and space translations, time-space dilatations with dynamical exponent $\mathpzc{z}=1$ and whose
Lie algebras contain conformal Lie algebras as sub-algebras. 
They act as} dynamical symmetries of the linear transport equation in $d$ spatial dimensions.
\BLAU{For $d=1$ spatial dimensions, meta-conformal transformations constitute new representations of the conformal Lie algebras,} 
while for $d\ne 1$ their algebraic structure is different.
Infinite-dimensional Lie algebras of meta-conformal transformations are explicitly constructed for $d=1$ and $d=2$
and they are shown to be isomorphic to the direct sum of either two or three centre-less Virasoro algebras, respectively.
\BLAU{The form of co-variant two-point correlators is derived. An application to the directed Glauber-Ising chain with
spatially long-ranged initial conditions is described.} 
\end{abstract}
\end{titlepage}

\setcounter{footnote}{0}

\begin{quote}
{\it ``Whenever you have to do with a structure-endowed entity, try to determine its group of automorphisms.
You can expect to gain a deep insight.''} \\
~\hfill H. Weyl, {\it Symmetry}, Princeton University Press (1952)
\end{quote}

%%%%%%%%%%%%%%%%%%%%%%%%%%%%%%%%%%%%%%%%%%%%%%%%%%%%%%%%%%%%%%%%%%%%%%%%%%%%%%%%%%%%%%%%%%%%%%%%%%%%%%%%%%%%%%%%%%%%%%%%%%%%%%%%%%%%%%%%
\section{Introduction}
%%%%%%%%%%%%%%%%%%%%%%%%%%%%%%%%%%%%%%%%%%%%%%%%%%%%%%%%%%%%%%%%%%%%%%%%%%%%%%%%%%%%%%%%%%%%%%%%%%%%%%%%%%%%%%%%%%%%%%%%%%%%%%%%%%%%%%%%

Conformal invariance has found many brilliant applications, for example to string theory and
high-energy physics \cite{Polchinski01}, or to two-dimensional phase transitions \cite{Belavin84,Francesco97,Henkel99,Rychkov17}
the quantum Hall effect \cite{Cappelli93,Hansson17}, or certain stochastic processes
\cite{Alcaraz12,Alcaraz13,Alcaraz15,Crampe16,Karevski17,Schuetz17}.
These applications are based on a geometric definition of conformal transformations, considered as local coordinate
transformations $\vec{r}\mapsto \vec{r}'=\vec{f}(\vec{r})$, of spatial coordinates $\vec{r}\in\mathbb{R}^2$
such that angles are kept unchanged.\footnote{See \cite{Rychkov17} and refs. therein for the considerable recent interest into the
case $\vec{r}\in\mathbb{R}^d$ with $d>2$.}
The Lie algebra of these transformations is naturally called the `{\it conformal Lie algebra}'.

\BLAU{{\bf 1.} In order to establish our notation, we briefly recall some basic facts, concentrating on $d=2$. 
Use complex light-cone coordinates $z=t+\II\mu r$ and $\bar{z}=t -\II\mu r$, 
where the `time' $t$ and the `space' $r$ label the two directions, and $\mu$ is an universal 
constant with the units of an inverse velocity. The Lie algebra generators read, for $n\in\mathbb{Z}$
\BEQ \label{1.1}
\ell_n = - z^{n+1} \partial_z - (n+1) \Delta z^n \:\; , \;\; 
\bar{\ell}_n = - \bar{z}^{n+1}\partial_{\bar{z}} - (n+1) \overline{\Delta} \bar{z}^n
\EEQ
where $\Delta=\demi(\delta-\II\gamma/\mu)$, $\overline{\Delta}=\demi(\delta+\II\gamma/\mu)$ 
are the conformal weights of the scaling operators 
on which these generators act. The generators (\ref{1.1})
obey the commutation relations of $\mathfrak{vect}(S^1)\oplus\mathfrak{vect}(S^1)$
\BEQ \label{1.2}
\left[ \ell_n, \ell_m \right] = (n-m) \ell_{n+m} \;\; , \;\;
\left[ \ell_n, \bar{\ell}_m \right] = 0  \;\; , \;\;
\left[ \bar{\ell}_n, \bar{\ell}_m \right] = (n-m) \bar{\ell}_{n+m} 
\EEQ
The maximal finite-dimensional Lie sub-algebra is 
$\mathfrak{conf}(2) := \left\langle \ell_{\pm 1,0}, \bar{\ell}_{\pm 1,0}\right\rangle 
\cong \mathfrak{sl}(2,\mathbb{R})\oplus\mathfrak{sl}(2,\mathbb{R})$. 
For what follows, we rather consider the generators $X_n := \ell_n + \bar{\ell}_n$ and 
$Y_n := \II\mu \bigl( \ell_n - \bar{\ell}_n\bigr)$ which in `time' and `space' coordinates read
\BEA
X_n &=& -\demi\left[ (t+\II\mu r)^{n+1} + (t-\II\mu r)^{n+1} \right]\partial_t 
        + \frac{\II}{2\mu} \left[ (t+\II\mu r)^{n+1} - (t-\II\mu r)^{n+1} \right]\partial_r \nonumber \\
    & & -\frac{n+1}{2} \delta \left[ (t+\II\mu r)^{n} + (t-\II\mu r)^{n} \right] 
        -\frac{n+1}{2} \frac{\gamma}{\II\mu} \left[ (t+\II\mu r)^{n} - (t-\II\mu r)^{n} \right]
        \nonumber \\
Y_n &=& -\frac{\II\mu}{2}\left[ (t+\II\mu r)^{n+1} - (t-\II\mu r)^{n+1} \right]\partial_t 
        - \demi \left[ (t+\II\mu r)^{n+1} + (t-\II\mu r)^{n+1} \right]\partial_r \nonumber \\
    & & -\frac{n+1}{2} \II\mu \delta \left[ (t+\II\mu r)^{n} - (t-\II\mu r)^{n} \right] 
        -\frac{n+1}{2} \gamma \left[ (t+\II\mu r)^{n} + (t-\II\mu r)^{n} \right]
\label{1.3}
\EEA
Herein, $\delta$ and $\gamma$ denote the scaling dimension and the (rescaled) spin of the respective scaling operator. 
The commutators (\ref{1.2}) are recast into 
\BEQ \label{1.4} 
\left[ X_n, X_m \right] = (n-m) X_{n+m} \;\; , \;\;
\left[ X_n, Y_m \right] = (n-m) Y_{n+m}  \;\; , \;\;
\left[ Y_n, Y_m \right] = -\mu\, (n-m) X_{n+m} 
\EEQ
These conformal transformations also act as dynamical symmetries of differential equations. 
{\em `Dynamical symmetry'} means throughout \cite{Niederer72} 
that the space of solutions of the equation $\hat{\mathscr{S}}\phi=0$ is invariant under the 
conformal transformations (\ref{1.1}). The most simple example is the
Laplace equation $\hat{\mathscr{S}}\phi=0$, where $\phi=\phi(z,\bar{z})=\vph(t,r)$ and
\BEQ \label{gl:Laplace}
\hat{\mathscr{S}} = 4\mu^2 \partial_z \partial_{\bar{z}} = \mu^2 \partial_t^2 + \partial_r^2
\EEQ
The dynamical conformal symmetry of (\ref{gl:Laplace}) follows from the commutators
\BEA
\left[ \hat{\mathscr{S}},\ell_n \right] &=& 
       - (n+1) z^n \hat{\mathscr{S}} - 4 n(n+1) \mu^2 \Delta z^{n-1}\partial_{\bar{z}} \nonumber \\
\left[ \hat{\mathscr{S}},\bar{\ell}_n \right] &=& 
       - (n+1) \bar{z}^n \hat{\mathscr{S}} - 4 n(n+1) \mu^2 \overline{\Delta} \bar{z}^{n-1}\partial_{z} 
\EEA 
and provided that $\Delta=\overline{\Delta}=0$.
Of course, the physical interest in conformal invariance comes from the multitude of systems,
beyond the Laplace equation and which are conformally invariant, as mentioned above. 
Finally, the requirement of co-variance under conformal transformations is sufficient to
fix certain $n$-point functions of the scaling operators $\phi_i(z_i,\bar{z}_i)$. 
For example, the two-point function $C(z,\bar{z}) = C(t,r)$ reads, up to normalisation
\BEA
C(z,\bar{z}) &=& \left\langle \phi_1(z,\bar{z}) \phi_2(0,0) \right\rangle 
\:=\: \delta_{\Delta_1,\Delta_2} \delta_{\overline{\Delta}_1,\overline{\Delta}_2}\: z^{-2\Delta_1} \bar{z}^{-2\overline{\Delta}_1} 
\nonumber \\
&=& \left\langle \vph_1(t,r) \vph_2(0,0) \right\rangle 
\:=\: \delta_{\delta_1,\delta_2} \delta_{\gamma_1,\gamma_2}\, \left( t^2 + \mu^2 r^2 \right)^{-\delta_1} 
\exp\left(-\frac{2\gamma_1}{\mu} \arctan\left( \mu \frac{r}{t}\right) \right) ~~~~
\label{1.6} 
\EEA}

\BLAU{{\bf 2.} 
Are there other groups of time-space transformations which can act as dynamical symmetries in certain physical situations~? 
In table~\ref{tab1}, several examples of infinite-dimensional Lie groups of time-space transformations are listed. 
The Schr\"odinger-Virasoro group \cite{Henkel94,Henkel03a,Unterberger12} is distinct from the conformal group in that the
dilatations are of the form $t\mapsto b^{\mathpzc{z}} t$ and $\vec{r}\mapsto b \vec{r}$, with the 
{\em dynamical exponent} $\mathpzc{z}=2$ for the Schr\"odinger group, in contrast to $\mathpzc{z}=1$ for the conformal group. 
Its maximal finite-dimensional subgroup 
is the {\em Schr\"odinger group} \cite{Jacobi1842,Lie1881,Niederer72,Burdet72,Hagen72,Jackiw72}, which acts as dynamical 
symmetry on the free diffusion/Schr\"odinger equation. Schr\"odinger-covariance predicts the form of response functions, as they
arise for example in phase-ordering kinetics, notably in non-equilibrium $2D$ and $3D$ Ising and Potts models, whose dynamics
are not described by free-field theories.} See \cite{Henkel10} for a review and \cite{Henkel17c} for a tutorial introduction.

%%++++++++++++++++++++++++++++++++++++++++++++++++++++++++++++++++++++++++++++++++++++++++++++++++++++++++++++++++++++++++++++++++++++%%
\begin{table}
\begin{center}\begin{tabular}{|l|lll|l|} \hline
group                    & \multicolumn{3}{l|}{coordinate changes}                                                  & co-variance \\
\hline
ortho-conformal $(1+1)D$ & $z'=f(z)$      & $\bar{z}'=\bar{z}$                        &                             & correlator  \\
                         & $z'=z$         & $\bar{z}'=\bar{f}(\bar{z})$               &                             & \\ \hline
Schr\"odinger-Virasoro   & $t'=b(t)$      & \multicolumn{2}{l|}{$\vec{r}'=\left(\D b(t)/\D t\right)^{1/2} \vec{r}$} & response \\
                         & $t'=t$         & $\vec{r}'=\vec{r}+\vec{a}(t)$             &                             & \\ 
                         & $t'=t$         & $\vec{r}'=\mathscr{R}(t)\vec{r}$          &                             & \\ \hline
conformal galilean       & $t'=b(t)$      & \multicolumn{2}{l|}{$\vec{r}'=\left(\D b(t)/\D t\right) \vec{r}$}       & correlator \\
                         & $t'=t$         & $\vec{r}'=\vec{r}+\vec{a}(t)$             &                             & \\ 
                         & $t'=t$         & $\vec{r}'=\mathscr{R}(t)\vec{r}$          &                             & \\ \hline
meta-conformal $1D$      & $u=f(u)$       & $\bar{u}'=\bar{u}$                        &                             & correlator \\
                         & $u'=u$         & $\bar{u}'=\bar{f}(\bar{u})$               &                             & \\ \hline
meta-conformal $2D$      & $\tau'=\tau$   & $w'=f(w)$                                 & $\bar{w}'=\bar{w}$          & \\
                         & $\tau'=\tau$   & $w'=w$                                    & $\bar{w}'=\bar{f}(\bar{w})$ & \\
                         & $\tau'=b(\tau)$ & $w'=w$                                   & $\bar{w}'=\bar{w}$          & \\ \hline
\end{tabular}\end{center}
\caption[tab1]{{\small
Examples of infinite-dimensional groups of time-space transformations, with the defining coordinate changes.
Herein, $f,\bar{f},b,\vec{a}$ are arbitrary differentiable (vector) functions of their argument and 
$\mathscr{R}(t)\in\mbox{\sl SO}(d)$ is a time-dependent rotation matrix.
Physical interpretations of the coordinates $(u,\bar{u})$ and $(\tau,w,\bar{w})$ of the $1D$ and $2D$ meta-conformal
transformations are listed in tables~\ref{tab2} and~\ref{tab3}.
The physical interpretation of the co-variant $n$-point functions as either correlators or responses is based on the extension of
the Cartan sub-algebra \cite{Henkel14a,Henkel15,Henkel16}.} \label{tab1}}
\end{table}
%%++++++++++++++++++++++++++++++++++++++++++++++++++++++++++++++++++++++++++++++++++++++++++++++++++++++++++++++++++++++++++++++++++++%%

\BLAU{If we concentrate on systems with a dynamical exponent $\mathpzc{z}=1$, can one find infinite-dimensional groups of
time-space transformations distinct from the conformal transformations reviewed above~? For the sake of a clear
conceptual distinction, those standard conformal transformations, generated from (\ref{1.1}) or (\ref{1.3}), 
will from now on be called {\em `ortho-conformal'}. It will turn out
that alternative sets of time-space transformations exist. In contrast to ortho-conformal transformations, these new transformations
are not angle-preserving, neither in a space made from time-space points $(t,\vec{r})\in\mathbb{R}^{1+d}$, 
nor in space with points $(\vec{r})\in\mathbb{R}^d$. 
On the other hand, their Lie algebras still contain
ortho-conformal Lie algebras as sub-algebras. We shall therefore call them 
{\it `meta-conformal transformations'} \cite{Henkel17a,Henkel17b}.}

{\bf 3.} In a two-dimensional time-space with points $(t,r)\in\mathbb{R}^2$, meta-conformal transformations have the
infinitesimal generators \cite{Henkel02}
\BEA
X_n   &=& -t^{n+1}\partial_t-\mu^{-1}[(t+\mu r)^{n+1}-t^{n+1}]\partial_r
          - (n+1)\frac{\gamma}{\mu}[(t+\mu r)^{n}-t^{n}] -(n+1)\delta t^n\nonumber\\
Y_{n} &=& -(t+\mu r)^{n+1}\partial_r- (n+1)\gamma (t+\mu r)^{n}
\label{infinivarconf}
\EEA
where $\delta,\gamma$ are constants and $\mu^{-1}$ is a constant universal velocity (`speed of sound' or `speed of light').
\BLAU{The generators $X_{-1}=-\partial_t$ and $Y_{-1}=-\partial_r$ of time- and space-translations, as well as the 
generator $X_0=-t\partial_t - r\partial_r -\delta$ of dilatations are the same as for ortho-conformal transformations (\ref{1.3}). 
The other generators are different and the generators (\ref{infinivarconf}) are in general not angle-preserving.} 
Their Lie algebra $\langle X_n, Y_n\rangle_{n\in \mathbb{Z}}$ obeys
\BEQ
[X_n,X_{m}] = (n-m)X_{n+m},\quad  [X_n,Y_{m}] = (n-m)Y_{n+m},\quad [Y_n,Y_{m}] = \mu (n-m)Y_{n+m}
\label{commutators}
\EEQ
The maximal finite-dimensional Lie sub-algebra is denoted $\mathfrak{meta}(1,1):=\left\langle X_{\pm 1,0}, Y_{\pm 1,0}\right\rangle$. 
Indeed, if $\mu\ne 0$, (\ref{commutators}) is isomorphic to the Lie algebra (\ref{1.4}). To see this, let
$X_n=\ell_n +\bar{\ell}_n$ and $Y_n=\mu\bar{\ell}_n$. This gives
\BEA
     \ell_n       & = & -t^{n+1}\left(\partial_t -\frac{1}{\mu}\partial_r\right) -(n+1)\left(\delta - \frac{\gamma}{\mu}\right)t^n
\nonumber\\
     \bar{\ell}_n & = & -\frac{1}{\mu}(t+\mu r)^{n+1}\partial_r                  -(n+1)\frac{\gamma}{\mu}(t+\mu r)^n.
\label{eq:ellconf}
\EEA
which again satisfy the commutators (\ref{1.2}). The  \BLAU{reduction of (\ref{eq:ellconf}) to the standard form (\ref{1.1})} 
in `complex' light-cone coordinates 
$z,\bar{z}$ is achieved by setting $z=t$ and $\bar{z}=t+\mu r$, and identifying the
conformal weights $\Delta=\delta-\gamma/\mu$ and $\overline{\Delta}=\gamma/\mu$. In $1+1$ time-space dimensions, 
the meta-conformal transformations (\ref{infinivarconf}) and the ortho-conformal transformations (\ref{1.4}) 
are two representations of the same conformal Lie algebra, see also table~\ref{tab1}.

The meta-conformal generators (\ref{infinivarconf}) are dynamical symmetries of the equation of motion
\BEQ \label{ineq1}
\hat{\mathscr{S}}\phi(t,r)=(-\mu\partial_t+\partial_r)\vph(t,r)=0.
\EEQ
Indeed, since (with $n\in\mathbb{Z}$)\index{conformal invariance}
\BEQ \label{dynsym}
{} [\hat{\mathscr{S}},X_n] = -(n+1) t^n \hat{\mathscr{S}} +n(n+1)\mu \left( \delta -\frac{\gamma}{\mu}\right)t^{n-1} \;\; , \;\;
{} [\hat{\mathscr{S}},Y_{n}] = 0
\EEQ
a solution $\vph$ of  (\ref{ineq1}) with scaling dimension
$\delta_{\vph}=\delta=\gamma/\mu$ is mapped onto another solution of (\ref{ineq1}).
Hence {\em the space of solutions of the equation (\ref{ineq1}) is meta-conformally invariant.} This is the analogue of the
ortho-conformal invariance of the $2D$ Laplace equation. \BLAU{This kind of equation of motion (\ref{ineq1}), with a directional bias, 
motivates to look for physical applications in the kinetics of spin systems with directed dynamics, as we shall do in section~5.} 

\BLAU{Meta-conformally co-variant two-point functions have the form \cite{Henkel10}, up to normalisation 
\BEQ \label{1.12} 
C(t,r) = \left\langle \vph_1(t,r) \vph_2(0,0) \right\rangle
= \delta_{\delta_1,\delta_2} \delta_{\gamma_1,\gamma_2}\: t^{-2\delta_1} 
\left( 1 + \frac{\gamma_1}{\mu} \frac{r}{t} \right)^{-2\gamma_1/\mu}
\EEQ}

\BLAU{{\bf 4.} 
In the limit $\mu\to 0$, for both ortho-conformal as well as for meta-conformal transformations, one can make a Lie algebra
contraction of (\ref{1.4}) or (\ref{commutators}).} The result is called {\em `conformal galilean algebra'} $\mbox{\sc cga}(1)$
\cite{Havas78} or {\em `{\sc bms}-algebra'} $\mathfrak{bms}_{3}$\cite{Bondi62}. 
\BLAU{Table~\ref{tab1} give the time-space transformations which follow from $\mbox{\sc cga}(d)$ for $d\geq 1$ 
(rotations by arbitrary time-dependent angles appear for $d\geq 2$). 
The generators of $\mbox{\sc cga}(1)$ can be read off by taking the limit $\mu\to 0$ in either 
(\ref{1.3}) or else in (\ref{infinivarconf}).\footnote{The conformal galilean generator 
$Y_0 = -t\partial_r -\gamma\in\mbox{\sc cga}(1)$ is distinct from the ordinary Galilei generator 
$Y_{1/2}=-t\partial_r - {\cal M} r\in \mathfrak{sch}(1)$ of the Schr\"odinger algebra, as these imply distinct transformations of 
the scaling operators.} Taking the limit $\mu\to 0$ in either (\ref{1.6}) or else (\ref{1.12}) 
gives the $\mbox{\sc cga}(1)$-covariant two-point function
\BEQ \label{1.13}
C(t,r) = \left\langle \vph_1(t,r) \vph_2(0,0) \right\rangle
= \delta_{\delta_1,\delta_2} \delta_{\gamma_1,\gamma_2}\: t^{-2\delta_1} 
\exp\left( - {2\gamma_1} \frac{r}{t} \right)
\EEQ}
The Lie algebra $\mbox{\sc cga}(d)$
is not isomorphic to the Schr\"odinger Lie algebra $\mathfrak{sch}(d)$ in $d$ dimensions \cite{Henkel03a,Duval09}.
\BLAU{An infinite-dimensional extension exists for all dimensions $d\geq 1$, see table~\ref{tab1}, and is distinct from the
Schr\"odinger-Virasoro group.} 
Applications arise in hydrodynamics \cite{Cherniha10,Zhang10} or in gravity, e.g.
\cite{Barnich07,Bagchi09,Bagchi10,Martelli10,Barnich13,Bagchi13b,Aizawa16},
and the bootstrap approach has been tried \cite{Martelli10,Bagchi17}.

\BLAU{Two-point functions such as (\ref{1.12},\ref{1.13}) display a singularity if $r/t$ becomes negative enough. This can be avoided
by (i) constructing an extension of the Cartan sub-algebra of meta-conformal transformations and (ii) applying the co-variance
conditions in an extended `dual' space, with respect to the `rapidities' $\gamma_i$ to considered as additional variables.} 
In $1D$, this gives the two-point function $C(t,r)=C_{12}(t,r)$ as \cite{Henkel16}
\BEQ \label{1.14}
C(t,r) = 
\delta_{\delta_1,\delta_2} \delta_{\gamma_1, \gamma_2}\:
|t|^{-2\delta_1} \left( 1 + \frac{\mu}{\gamma_1} \left| \frac{\gamma_1 r}{t} \right| \right)^{-2\gamma_1/\mu} 
\stackrel{\mu\to 0}{\xrightarrow{\hspace*{6mm}}}  \delta_{\delta_1,\delta_2} \delta_{\gamma_1, \gamma_2}\:
|t|^{-2\delta_1}  \exp\left( -\left| \frac{2\gamma_1 r}{t} \right|\right) 
\EEQ
One has the symmetry $C_{12}(t,r)=C_{21}(-t,-r)$ under
permutation of the scaling operators \BLAU{$\phi_{1}$ and $\phi_{2}$}, expected for a correlator. 
This is analogous to ortho-conformal
invariance.\footnote{For the Schr\"odinger group, an analogous construction shows that the two-point functions are response
functions \cite{Henkel03a,Henkel15,Henkel15a,Henkel16}. The scaling form (\ref{1.14}) of the meta-conformal correlator
is the same as the special case $\mathpzc{z}=1$ for the {\em conformally co-variant} two-time 
{\em response function} $G(t,r)$ \cite[eq. (3.10)]{Cardy85}.}

We mention further examples of physical systems with dynamical exponent $\mathpzc{z}=1$. 
First, the dynamical symmetries of the Jeans-Vlassov equation 
\cite{Jeans1915,Vlasov38,Henon82,Mo10,Vilani14,Campa09,Campa14,Elskens14,Pegoraro15}
in one space dimension are given by a representation of (\ref{commutators}),
distinct from (\ref{infinivarconf}) \cite{Stoimenov15}. 
Second, the non-equilibrium dynamics of \BLAU{open quantum systems after a quantum quench} generically
has $\mathpzc{z}=1$, related to ballistic spreading of signals, see \cite{Calabrese07,Calabrese16,Dutta15}
and this apparently holds both for quenches in the vicinity of the quantum critical point
\cite{Delfino17} as well as for deep quenches into the two-phase coexistence region \cite{Maraga15,Wald17}.
%%MH je ne crois plus que la phrase suivante soit vraie (9.6.19)
%%The available examples suggest that the
%%value $\mathpzc{z}=1$ should be robust with respect to the change from closed to open quantum systems.
Third, effective equations of motion of the form
(\ref{ineq1}) arise in recent studies of the generalised hydrodynamics
required for the description of strongly interacting non-equilibrium
quantum systems \cite{Bertini16,Castro16,Doyon17,Caux17,Piroli17,Dubail16}. \BLAU{Forth, we shall consider in section~5
the non-equilibrium relaxational dynamics in directed spin systems, such as the directed Glauber-Ising model 
\cite{Godreche11,Godreche15a,Godreche15b}.} 

\BLAU{{\bf 5.} Can one find meta-conformal transformations in $d\geq 2$ spatial dimensions~? We shall require 
that time- and space-translations, as well as dilatations with $\mathpzc{z}=1$, 
are kept in their form known from ortho-conformal invariance. 
Table~\ref{tab1} shows several examples of infinite-dimensional time-space transformations groups and how
meta-conformal transformations in $d=1$ or $d=2$ constructed in this work compare with other known 
examples.\footnote{The $2D$ meta-conformal case also arises from a systematic extension of L\'evy-Leblond's 
Carroll group in $(1+1)D$, where it is called the ``conformal $k=\infty$ Carroll Lie group'' \cite{Duval14a}.} 
Tables~\ref{tab2} and~\ref{tab3} below give the
physical interpretations of the formal abstract coordinates $(u,\bar{u})$ or $(\tau,w,\bar{w})$. used in table~\ref{tab1}. 
In this way, the analogies and differences 
between these distinct groups become apparent, notably concerning the transformation of the
spatial coordinates.\footnote{They differ also from Cardy's proposal
$(t,\vec{r})\mapsto (t',\vec{r}')=(b(\vec{r})^{\mathpzc{z}} t,b(\vec{r})\vec{r})$ \cite{Cardy85}.}
Only the ortho-conformal transformation include rotations between the `time' and `space' coordinates.}

This work is organised as follows. 
In section~2, a generalisation of the representation (\ref{infinivarconf}) of $1D$ meta-conformal transformations will
be presented. We shall give a geometrical interpretation of several types of meta-conformal
transformations. This allows to formulate an ansatz for the $d$-dimensional construction which is
used in section~3 to find the generic form of the generators \BLAU{of the Lie algebra of meta-conformal transformations, 
to be denoted by $\mathfrak{meta}(1,d)$.} 
Particular attention will be devoted to construct the terms which will
describe how primary scaling operators will transform under meta-conformal transformations.
In section~4 we shall concentrate on the special case of
$d=2$ dimensions, where stronger results are found.
First, we identify two distinct meta-conformal representations which are distinguished by different sets of physical
coordinates, as listed in table~\ref{tab3}. Second, while for $d>2$ this only gives a finite-dimensional
Lie algebra, we shall see for $d=2$ an infinite-dimensional extension exists
which is isomorphic to the direct sum of {\em three} Virasoro algebras
(without central charge).\footnote{The same algebra of dynamical symmetries also arises for diffusion-limited erosion
in $1D$ \cite{Henkel17a,Henkel17b}.}
The corresponding finite (group) transformations are indicated in table~\ref{tab1}.
The time-dependent transformations might be used to generate the temporal evolution of the physical system.
Indeed, the co-variant
two-point function is explicitly seen to describe the relaxation towards an  ortho-conformally two-point function,
which reflects the meta-conformal aspects in this Lie group.
\BLAU{Section~5 describes the application to the non-equilibrium relaxation behaviour of the directed Glauber-Ising chain, 
in the case of spatially longed-ranged initial conditions.} 
We conclude in section~6.

\newpage
%%%%%%%%%%%%%%%%%%%%%%%%%%%%%%%%%%%%%%%%%%%%%%%%%%%%%%%%%%%%%%%%%%%%%%%%%%%%%%%%%%%%%%%%%%%%%%%%%%%%%%%%%%%%%%%%%%%%%%%%%%%%%%%
\section{Meta-conformal algebras: general remarks}
%%%%%%%%%%%%%%%%%%%%%%%%%%%%%%%%%%%%%%%%%%%%%%%%%%%%%%%%%%%%%%%%%%%%%%%%%%%%%%%%%%%%%%%%%%%%%%%%%%%%%%%%%%%%%%%%%%%%%%%%%%%%%%%

\subsection{A generalisation of the one-dimensional case}

We begin by reconsidering the dynamical symmetries of eq.~(\ref{ineq1}), re-written in the form
\BEQ
\hat{\mathscr{S}}\vph(t,r)=(\partial_t+c\partial_r)\vph(t,r)=0\label{ineq2}
\EEQ
\BLAU{where $c$ is a constant. 
For what follows, we need a generalisation of the meta-conformal representation (\ref{infinivarconf}). 
By assumption, both time- and space-translations $X_{-1}$, $Y_{-1}$, 
as well as the dilatations $X_0$, retain their form given in (\ref{infinivarconf}).}
However, the explicit generators $X_1,Y_{0,1}$ of the finite-dimensional sub-algebra
$\mathfrak{meta}(1,1):=\langle X_n, Y_n\rangle_{n\in\{\pm 1,0\}}$
\BLAU{admit in general the following form}, with the constants $\alpha,\beta$ \cite{Stoimenov15}:
\BEA
X_1   & = & -\left(t^2+\alpha r^{2}\right)\partial_t 
            - \left(2t r +\beta r^{2}\right)\partial_r-2\delta t - 2\gamma r,\nonumber\\
Y_{0} & = & -\alpha r\partial_t-\left(t + \beta r\right)\partial_r -\gamma \nonumber\\
Y_{1} & = & -\alpha\left(2t r+\beta r^{2}\right)\partial_t
\label{extendsymmetries} \\
 && - \left(t^2+2\beta t r +(\alpha+\beta^2)r^{2}\right)\partial_r 
    - 2\gamma t -2\left(\alpha \delta+\beta\gamma\right)r. \nonumber
\EEA
For $n,m,\in \{0,\pm 1\}$ they satisfy the following commutation relations
\BEA
     &&[X_n, X_{m}]=(n-m)X_{n+m} \;\; ,\;\; [X_n, Y_m] = (n-m)Y_{n+m}\nonumber\\
     && [Y_n, Y_{m}] = (n-m)\left(\alpha X_{n+m} + \beta Y_{n+m}\right).
     \label{metaconformal}
\EEA
With respect to the meta-conformal generators (\ref{infinivarconf}), the new feature is the constant $\alpha\ne 0$. 

Furthermore, \BLAU{if we make the choice} \cite{Stoimenov15}
\BEQ \label{eq2.4}
\alpha=\frac{1+\beta c}{c^2} \;\; , \;\;
\delta=-\gamma c
\EEQ
then the generators (\ref{extendsymmetries}) are indeed dynamical meta-conformal symmetries of the  $1D$ 
\BLAU{transport equation}~(\ref{ineq2}). This follows from the commutators
\BEA
&& [\hat{\mathscr{S}}, X_1] = -2\left(t+\frac{1+\beta c}{c^2}cr\right)\hat{\mathscr{S}}\nonumber\\
&& [\hat{\mathscr{S}}, Y_0] = -\frac{1+\beta c}{c^2}\,c \hat{\mathscr{S}} \\
&& [\hat{\mathscr{S}}, Y_1] = -2\frac{1+\beta c}{c^2}\left(ct+(1+\beta c)r\right)\hat{\mathscr{S}}\nonumber
\EEA
\BLAU{Hence the solution space of $\hat{\mathscr{S}}\vph=0$ is invariant under the representation (\ref{extendsymmetries}).}

While commutators (\ref{metaconformal}),(\ref{commutators}) look different, the Lie algebra
$\langle X_n, Y_n\rangle_{n\in \{0,\pm 1\}} \cong \mathfrak{sl}(2,\mathbb{R})\oplus\mathfrak{sl}(2,\mathbb{R})$
is isomorphic to the ortho-conformal algebra $\mathfrak{conf}(2)$ \cite[Prop. 1]{Stoimenov15} 
and hence also to (\ref{commutators}). We want to find this isomorphism explicitly.

First, the case $\alpha=0$ reduces the algebra to the usual form
(\ref{commutators}). Then the choices $c=-1/\beta$ and $\beta=\mu$ make equations (\ref{ineq1}, \ref{ineq2}) co\"{\i}ncide.

Second, for $\alpha\ne 0$ there is no obvious relation between $c$ and $\beta$. Define new generators 
\BEQ
{\cal Y}_n := a X_n+Y_n\label{isomorph1} \;\; ; \;\; \mbox{\rm ~~for $n\in\{ 0, \pm 1\}$} 
\EEQ
Using (\ref{metaconformal}), it is easily seen that
\BEQ
[{\cal Y}_n, {\cal Y}_m]=(2a+\beta)(n-m){\cal Y}_{n+m}\nonumber
\EEQ
where $a$ must satisfy the quadratic equation $a^2+\beta a -\alpha=0$. 
The solutions are $a_{\pm}= (-\beta \pm \sqrt{\beta^2+4\alpha})/2$.
Rescaling the generators $Y_n$, one may effectively rescale one of the constants $\alpha,\beta$ as desired;
we shall take $\alpha=-\frac{2}{9}\beta^2$
in what follows.\footnote{This choice is motivated from the $d$-dimensional case with $d\geq 2$, see sections~3 and~4.}
Then we have the two cases $a_{-}=-\frac{2}{3}\beta$ and $a_{+}=-\frac{1}{3}\beta$.\\

\noindent\underline{\bf Case A}: \underline{$a=a_{-}=-\frac{2}{3}\beta$.}
Combining eq.~(\ref{eq2.4}) with $\mu=-1/c$ fixes $\mu =-\beta/3$. We have
\BEQ\label{defcalY}
{\cal Y}^{(A)}_n=-\frac{2}{3}\beta X_n+Y_n,
\EEQ
and recover the algebra (\ref{commutators}) in the finite-dimensional case
\BEQ\label{minisomorph}
[X_n, {\cal Y}^{(A)}_m] =(n-m){\cal Y}^{(A)}_{n+m},\quad 
[{\cal Y}^{(A)}_n,{\cal Y}^{(A)}_m]=-\frac{\beta}{3}(n-m){\cal Y}^{(A)}_{n+m}.
\EEQ
In addition, from this representation, see (\ref{extendsymmetries}),
of the algebra $\left\langle X_{0,\pm 1}, {\cal Y}^{(A)}_{0,\pm 1}\right\rangle$ 
with commutators (\ref{metaconformal},\ref{minisomorph}),
an  infinite-dimensional extension can be found. To do so, we first define
\BEQ \label{defAnA}
A^{(A)}_n=X_n+ \frac{3}{\beta} {\cal Y}^{(A)}_n  =-X_n+\frac{3}{\beta}Y_n
\EEQ
with the following simplified commutators
\BEQ \label{sl2rmconfA}
[A^{(A)}_n, A^{(A)}_m]=(n-m)A^{(A)}_{n+m},\quad [{\cal Y}^{(A)}_n, {\cal Y}^{(A)}_m]=-\frac{\beta}{3}(n-m){\cal Y}^{(A)}_{n+m},
\quad [A^{(A)}_n, {\cal Y}^{(A)}_m]=0.
\EEQ
The explicit representation for all $n\in\mathbb{Z}$ will be given below.

\noindent\underline{\bf Case B}: \underline{$a=a_{+}=-\frac{\beta}{3}$.} We now have
\BEQ\label{BdefcalY}
{\cal Y}^{(B)}_n=-\frac{\beta}{3}X_n+Y_n \;\; , \;\; A^{(B)}_n=X_n-\frac{3}{\beta}{\cal Y}^{(B)}_n 
\EEQ
\BLAU{However, it is unnecessary to reproduce the commutators, since the cases A and B are not independent. 
Rather, we have (for $\beta\ne 0,\infty$ and $\alpha\ne 0$)}
\BEQ
A^{(B)}_n        = 2X_n-\frac{3}{\beta}Y_n = -\frac{3}{\beta}{\cal Y}^{(A)}_n \;\; , \;\; 
{\cal Y}^{(B)}_n = -\frac{\beta}{3}X_n+Y_n = \frac{\beta}{3}A^{(A)}_n.
\label{splittingAB}
\EEQ

%%+++++++++++++++++++++++++++++++++++++++++++++++++++++++++++++++++++++++++++++++++++++++++++++++++++++++++++++++++++++++++++%%
\begin{table}
\begin{center}\begin{tabular}{|ll|ll|ll|} \hline
transformation                      &                & \multicolumn{1}{c}{$u$} & \multicolumn{1}{c|}{$\bar{u}$}  
                                    & \multicolumn{1}{c}{$\Delta$}             & \multicolumn{1}{c|}{$\overline{\Delta}$} \\ \hline
ortho-conformal $(1+1)D$            &                & $z=t+\II r$             & $\bar{z}=t-\II r$    
                                    & $\demi\bigl(\delta-\frac{\II\gamma}{\mu}\bigr)$ 
                                    & $\demi\bigl(\delta+\frac{\II\gamma}{\mu} \bigr)$\\[0.12truecm]
meta-conformal $1D$ \hfill {\sc i}  & $\alpha=0$     & $t$                     & $\rho=t+\mu r$  
                                    & $\delta-\frac{\gamma}{\mu}$              & $\frac{\gamma}{\mu}$ \\[0.12truecm]
meta-conformal $1D$ \hfill {\sc ii} & $\alpha\ne 0$  & $t+\frac{2\beta}{3}\,r$ & $t+\frac{\beta}{3}\,r$   
                                    & $\frac{3\gamma}{\beta}-\delta$           & $2\delta -\frac{3\gamma}{\beta}$ \\[0.12truecm] \hline
\end{tabular}\end{center}
\caption[tab2]{\small Possible choices for the `complex' light-cone coordinates $u,\bar{u}$ of the
conformal generators $\ell_n=-u^{n+1}\partial_u =(n+1)\Delta u^n$ and
$\bar{\ell}_n -\bar{u}^{n+1}\partial_{\bar{u}} -(n+1)\overline{\Delta}\: \bar{u}^n$. The meta-conformal representations are
eqs.~(\ref{eq:ellconf},\ref{2infinitmconf}) for $\alpha=0$ and $\alpha\ne0$, respectively. 
\BLAU{The resulting conformal weights
$\Delta,\overline{\Delta}$ are also indicated. In case {\sc ii}, $\mu=-\beta/3$ and the scaling 
$\alpha=-\frac{2}{9}\beta^2=-2\mu^2$ was used.} 
\label{tab2}}
\end{table}
%%+++++++++++++++++++++++++++++++++++++++++++++++++++++++++++++++++++++++++++++++++++++++++++++++++++++++++++++++++++++++++++%%

\BLAU{Concentrating on case A, and letting $A_n=A_n^{(A)}$ and ${\cal Y}_n=-{\cal Y}^{(A)}_n$,
we have the following infinite-dimensional representation of 
meta-conformal transformations, for the chosen rescaling $\alpha=-\frac{2}{9}\beta^2\ne 0$}
\BEA
A_n & = & -\left(t+\frac{2}{3}\beta r\right)^{n+1}\left(\frac{3}{\beta}\partial_r-\partial_t\right)
         -(n+1)\left(\frac{3}{\beta}\gamma-\delta\right) \left(t+\frac{2}{3}\beta r\right)^n
\nonumber\\
{\cal Y}_n & = &  -\left(t+\frac{\beta}{3}r\right)^{n+1}\left(\frac{2}{3}\beta\partial_t-\partial_r\right)
                -(n+1)\left(\frac{2}{3}\beta\delta-\gamma\right) \left(t+\frac{\beta}{3}r\right)^n
\label{2infinitmconf}
\EEA
with the commutation relations, for $n,m\in\mathbb{Z}$
\BEQ \label{29sl2rmconf}
[A_n, A_m]=(n-m)A_{n+m},\quad [{\cal Y}_n, {\cal Y}_m]=\frac{\beta}{3}(n-m){\cal Y}_{n+m}, \quad [A_n, {\cal Y}_m]=0.
\EEQ
\BLAU{In particular, for $n=\pm 1,0$ the generators (\ref{extendsymmetries}) are reproduced.}
The generators $A_n,{\cal Y}_n$ are the analogues of the generators $\ell_n,\bar{\ell}_n$ from
the representation (\ref{eq:ellconf}) of $1D$ meta-conformal
invariance, see table~\ref{tab2}. Indeed, with the light-cone coordinates
\BEQ
u = t +\frac{2\beta}{3} r \;\; , \;\; \bar{u} = t +\frac{\beta}{3} r
\EEQ
the generators (\ref{2infinitmconf}) reduce to the usual ortho-conformal form \cite{Belavin84}
\BD
\ell_n \leftrightarrow A_n = -u^{n+1}\partial_u - (n+1)\Delta\, u^n \;\; , \;\;
\bar{\ell}_n \leftrightarrow 
\frac{3}{\beta}{\cal Y}_n = - \bar{u}^{n+1} \partial_{\bar{u}} -(n+1)\overline{\Delta}\, \bar{u}^n 
\ED
with the conformal weights $\Delta = \frac{3\gamma}{\beta}-\delta$ and $\overline{\Delta} = 2\delta-\frac{3\gamma}{\beta}$.

%%====================================================================================================================================%%
\begin{figure}[tb]
\begin{center}
\includegraphics[width=.8\hsize]{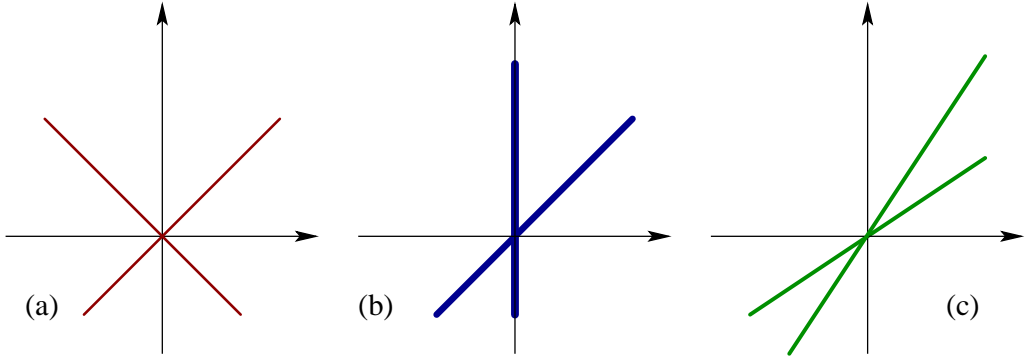}
\end{center}
\caption{\small Comparison of the light-cone coordinates $u,\bar{u}$ for
(a) $2D$ ortho-conformal transformations, (b) $1D$ meta-conformal transformations
with $\alpha=0$ and (c) $1D$ meta-conformal transformations with $\alpha\ne0$.  \label{figcoor}}
\end{figure}
%%===================================================================================================================================%%

\BLAU{Summarising, we have have found the distinct kinds of time-space transformations which arise from the conformal algebra
in $(1+1)D$, under the assumption stated above.} 

\noindent
\BLAU{{\bf Proposition 1:} {\em For $1+1$ time-space dimensions, the distinct representations as time-space transformations 
of the conformal algebra (\ref{1.2}) 
are listed in table~\ref{tab2}. The choice of orthogonal coordinates $u,\bar{u}$ corresponds to ortho-conformal transformations, 
while meta-conformal transformations are found if non-orthogonal coordinates $u,\bar{u}$ are used.}}

The different physical interpretations of the light-cone coordinates $u,\bar{u}$ are illustrated in figure~\ref{figcoor}.
Clearly, the `natural' coordinates of $1D$ meta-conformal transformations do {\em not} correspond to orthogonal coordinates,
\BLAU{while ortho-conformal transformations are obtained for orthogonal coordinates.}

\subsection{Finite $1D$ meta-conformal transformations}

\BLAU{A more clear geometric picture of the meta-conformal transformations of table~\ref{tab2} 
can be found by constructing} the Lie series
$F_Y(\eps,t,r) = e^{\eps Y_{m}} F(0,t,r)$ and $F_X(\eps,t,r) = e^{\eps X_{n}} F(0,t,r)$ of the corresponding finite transformations.
If we use \BLAU{the representation} (\ref{infinivarconf}),
they are given as the solutions of the two initial-value problems (herein, $\alpha=0$)
\begin{subequations} \label{XYfinit}
\begin{align}
& \Bigl( \partial_{\eps} + (t+\mu r)^{m+1}\partial_r + (m+1)\gamma(t+\mu r)^m\Bigr) F_Y(\eps,t,r) = 0
\label{finitYm}\\
& \Bigl( \partial_{\eps} +t^{n+1}\partial_t+\mu^{-1}
\left[(t+\mu r)^{n+1}-t^{n+1}\right]\partial_r+(n+1)\left(\delta t^n+\frac{\gamma}{\mu}\left[(t+\mu r)^n-t^n\right]\right)\Bigr)
F_X(\eps,t,r) = 0
\label{finitXn}
\end{align}
\end{subequations}
subject to the initial conditions $F_X(0,t,r)=F_Y(0,t,r)=\vph(t,r)$. If we work with the variable $\rho=t+\mu r$ instead of $r$,
we write the initial condition $F_X(0,t,\rho)=F_Y(0,t,\rho)=\phi(t,\rho):=\vph(t,r)$ and find, \BLAU{with the conformal weights
$\Delta$, $\overline{\Delta}$ taken from table~\ref{tab2}} 
\begin{subequations} \label{finit-1D-A}
\begin{align}
Y_m:&\quad \phi'(t,\rho)  =  \left(\frac{\D \rho'}{\D\rho}\right)^{\overline{\Delta}}\phi(t',\rho')\;\;\hspace{1.65truecm};\;\;
         t'=t\;\;,\;\; \rho'  =  a(\rho) \\
X_n:&\quad \phi'(t,\rho) =  \left(\frac{\D t'}{\D t}\right)^{\Delta}
                            \left(\frac{\D \rho'}{\D\rho}\right)^{\overline{\Delta}} \phi\left(t',\rho'\right)\;\;;\;\;
t'=b(t) \;\;,\;\; \rho'=b(\rho)
\end{align}
\end{subequations}
where $a=a(\rho)$ and $b=b(t)$ are arbitrary differentiable functions. The transformation of $r$ as generated by $X_n$ reads 
\BEQ
r'=\frac{1}{\mu}[b(t+\mu r)-b(t)].\label{rtransf}
\EEQ
Re-expanding $b(t)=t-\eps t^{n+1}$ and $a(\rho)=\rho-\eps \rho^{m+1}$ reproduces the differential
equations (\ref{XYfinit}) for the Lie series. 
Alternatively, \BLAU{if $\alpha\ne 0$}, we can use the representation (\ref{2infinitmconf}), \BLAU{and work with the
variables $u$ and $\bar{u}$, as well as with the conformal weights $\Delta$ and $\overline{\Delta}$, 
as given in table~\ref{tab2}} and find
\begin{subequations} \label{finit-1D-B}
\begin{align}
{\cal Y}_m:&\quad \phi'(u,\bar{u})  =  \left(\frac{\D \bar{u}'}{\D \bar{u}}\right)^{\overline{\Delta}}\phi(u',\bar{u})\;\; 
\hspace{0.25truecm};\;\;
         u'=u\;\;,\;\; \bar{u}'  =  \bar{f}(\bar{u}) \\
A_n:&\quad \phi'(u,\bar{u}) =  \left(\frac{\D u'}{\D u}\right)^{\Delta} \phi\left(u',\bar{u}'\right)\;\; ;\;\;
         u'=f(u) \;\;,\;\; \bar{u}'=\bar{u}
\end{align}
\end{subequations}
and where $f=f(u)$ and $\bar{f}=\bar{f}(\bar{u})$ are arbitrary differentiable functions. 
\BLAU{This is the statement already contained in table~\ref{tab1}, which remains valid for $\alpha=0$ and $\alpha\ne 0$.} 
%As before, the explicit transformation of time $t$ and space $r$ can be reconstructed.

\BLAU{Eqs.~(\ref{finit-1D-A},\ref{finit-1D-B}) give the global form of the $1D$
meta-conformal transformations and show how a {\em primary} meta-conformal scaling operator should be defined, generalising
the concept from the case of orthogonal coordinates studied in \cite{Belavin84} to non-orthogonal coordinates.} 
In this work, we shall concentrate on finding 
new meta-conformal transformations and we shall leave the construction of the full conformal field-theory
based on (\ref{finit-1D-A},\ref{finit-1D-B}) to future work.

\subsection{Ansatz for the $d$-dimensional case}

Meta-conformal transformations are analogues of the conformal algebra $\mathfrak{conf}(d)$ and  are sought as dynamical symmetries
of a ballistic transport equation, with a constant $\vec{c}\in\mathbb{R}^d$, 
\BEQ  \label{Boltzmannd}
\hat{\mathscr{S}}\phi(t,\vec{r})=(\partial_t+\vec{c}\cdot \vec{\partial}_{\vec{r}})\phi(t,\vec{r})=0
\EEQ
This naturally generalises eq.~(\ref{ineq1}).
\BLAU{Our construction starts from two axioms:} 

\noindent 
{\bf (i)} The generators of translations and \BLAU{time-space dilatations read in} $d$ dimensions
\begin{subequations} \label{gentriviaux}
\begin{align}
     X_{-1}   &=  -\partial_t                                         \label{timetranslations}\\
     Y^j_{-1} &=  -\partial_{r_j}\;\; , \quad j\in \{1,\ldots, d\}    \label{spacetranslations}\\
     X_0      &=  -t\partial_t-\vec{r}\cdot\partial_{\vec{r}}-\delta  \label{dynscaling}
\end{align}
\end{subequations}
where $\delta$ stands for a scaling dimension. If $d\leq 3$, we shall also write $j\in\{x,y,z\}$.

\noindent
{\bf (ii)} Specifying $X_1$ fixes all further generators. We make the ansatz
\BEA
X_1 & := & -(t^2+\alpha\vec{r}^2)\partial_t-2t\vec{r}\cdot\partial_{\vec{r}}-p(\vec{r}\cdot\vec{r})\vec{\beta}\cdot\partial_{\vec{r}}-
           (1-p)(\vec{\beta}\cdot\vec{r})\vec{r}\cdot\partial_{\vec{r}}-2\delta t\nonumber\\
    &    & -\vec{B}(\vec{r}, \vec{\beta}, \vec{\gamma})\cdot\partial_{\vec{\gamma}}
           -k\vec{\gamma}\cdot\vec{r}  \label{X1ddim}
\EEA
where $\alpha$, $p$ and $k$ are scalars, $\vec{\beta}$, $\vec{\gamma}$ are vectors and
the vector $\vec{B}$ depends on its arguments. All these must be found self-consistently from the algebra we are going to construct.

By construction, $[X_n,X_m]=(n-m)X_{n+m}$ is obeyed for $n,m\in\{\pm 1,0\}$.
All further generators of the Lie algebra will be obtained from repeated commutators of $X_1$ with $X_{-1}$ and $Y_{-1}^j$,
\BLAU{using $\left[ X_1, Y_m^j\right]=(1-m)Y_{m+1}^j$}.
The form (\ref{X1ddim}) of the generator $X_1$ is motivated as follows.
\begin{itemize}
\item For $d=1$ dimension, one should reproduce $X_1$ in eq.~(\ref{extendsymmetries}).
The $1D$ generator contains a term $-\beta r^2\partial_r$, which for $d>1$ leads
to two distinct contributions, as specified in (\ref{X1ddim}).
\item $X_1$ should be rotation-invariant, that is it should commute with the generators $R_{ij}$ of spatial rotations.
However, for the `natural' choice $R_{ij}= r_i \partial_{r_j} - r_j \partial_{r_i}$, the
invariance condition $\left[ X_1, R_{ij}\right]=0$ does not hold, not even in the special case  $\vec{B}=\vec{\gamma}=0$.
Therefore, spatial rotations should also include rotations of the vectors $\vec{\beta}$ and $\vec{\gamma}$. The
rotation generator becomes
\BEQ \label{newrotations}
\bar{R}_{ij} = \left(r_i \partial_{r_j} - r_j \partial_{r_i}\right) +
\vep_{\gamma}\left(\gamma_i \partial_{\gamma_j} - \gamma_j \partial_{\gamma_i}\right)+
\vep_{\beta}\left(\beta_i \partial_{\beta_j} - \beta_j \partial_{\beta_i}\right)
\EEQ
where the signatures $\vep_{\gamma},\vep_{\beta}=\pm 1$ allow for a different sense of rotation of
$\vec{\beta}$ or $\vec{\gamma}$ than of the spatial coordinates $\vec{r}$.
Furthermore, we should allow for the possibility $\vec{B}\ne \vec{0}$.
In addition, from the commutation relation of the one-dimensional case (\ref{commutators}),
especially $[[X_1, Y^j_{-1}], Y_{-1}^j]\sim Y_{-1}^j$, and (\ref{spacetranslations}),
it follows that $\vec{B}$ can be at most linear in $\vec{r}$.
\end{itemize}
Additional restrictions on the form of $X_1$ come from the requirement that it should act
as a dynamical symmetry of eq.~(\ref{Boltzmannd}).
By `dynamical symmetry' we mean the following required commutator \cite{Niederer72}
\BEQ \label{condsimX1}
[\hat{\mathscr{S}}, X_1]\phi=\lambda(t,\vec{r})\hat{\mathscr{S}}\phi.
\EEQ
which implies that the space of solutions of $\hat{\mathscr{S}}\phi=0$ is invariant under the action of $X_1$ (eventually after fixing
one or several scaling dimensions of $\phi$ to certain values).
As we shall see, this requirement leads to new relations between $\alpha,p$ and $\vec{\beta}$.

\noindent \underline{\bf Example:} The two vectors $\vec{\beta}$ and $\vec{c}$ span a two-dimensional space.
By rotation-invariance, it is therefore enough to consider the case $d=2$, since
any higher-dimensional situation can be reduced to the present case.
Let $\vec{B}=\vec{\gamma}=\vec{0}$. From (\ref{X1ddim},\ref{condsimX1}) it follows that $\delta=0$ and
\begin{subequations} \label{syst}
\begin{align}
1+\beta_xc_x+\frac{1-p}{2}\beta_yc_y  &= \alpha c_x^2  \label{sys1}\\
p\beta_xc_y+\frac{1-p}{2}\beta_yc_x   &= \alpha c_xc_y \label{sys2}\\
1+\frac{1-p}{2}\beta_xc_x+\beta_yc_y  &= \alpha c_y^2  \label{sys4}\\
p\beta_yc_x+\frac{1-p}{2}\beta_xc_y   &= \alpha c_xc_y \label{sys5}
\end{align}
\end{subequations}
We look for a solution of the above system for $\vec{\beta}\ne \vec{0}$. Straightforward calculations show:
\begin{enumerate}
\item The case $p=1$ leads to contradictions between some of the equations in the system (\ref{syst}).
Then the generator $X_1$ cannot be a symmetry. 
\item For $p\ne 1$, we have the following solution of the system (\ref{syst})
\begin{subequations} \label{2.30}
\begin{align}
c_j      & =  \frac{2}{p-1}\frac{\beta_j}{\vec{\beta}^2}\;\; , \quad j=x,y \label{eqsym}   \\
  \alpha & =  \frac{1}{4}(p+1)(p-1)\:\vec{\beta}^2                         \label{detalpha}
\end{align}
\end{subequations}
Hence, the condition (\ref{condsimX1}) is satisfied, with $\lambda(t,\vec{r})=-2t-(p+1)(\vec{\beta}\cdot\vec{r})$.
In contrast with the $1D$ case, $\alpha$ is fixed by (\ref{detalpha}) in terms of $\vec{\beta}$.
In particular, $\alpha=0$ is only possible for $p=-1$. The solution (\ref{2.30}) holds true for all dimensions $d>1$.

Eq.~(\ref{eqsym}) shows that $\vec{c}$ and $\vec{\beta}$ are collinear.
Calculations are simplified by choosing the orientation of the
coordinate axes such that only $\beta_1=\beta_x\ne 0$ and $\beta_j=0$ for all $j\geq 2$.
\item For $d=1$, only eq.~(\ref{sys1}) remains, which is equivalent to (\ref{eq2.4}).
Hence the structure of the $1D$ meta-conformal algebra
is distinct from the one in any other dimension $d>1$.
\end{enumerate}
In general, $\vec{B}\ne\vec{0}$ depends linearly on $\vec{r}$. Then the sought symmetries generated by $X_1$ can
become {\em conditional symmetries}, that is some auxiliary conditions on the field 
$\vph=\vph(t,\vec{r},\vec{\beta},\vec{\gamma})$
must be imposed, see \cite{Bluman69,Fushchych88,Fushchych93,Cherniha17} and references therein.
\BLAU{We shall come back to this} at the end of section~3.

%%%%%%%%%%%%%%%%%%%%%%%%%%%%%%%%%%%%%%%%%%%%%%%%%%%%%%%%%%%%%%%%%%%%%%%%%%%%%%%%%%%%%%%%%%%%%%%%%%%%%%%%%%%%%%%%%%%%%%%%%%%%%%%
\section{Meta-conformal algebra in $d>1$ spatial dimensions}
%%%%%%%%%%%%%%%%%%%%%%%%%%%%%%%%%%%%%%%%%%%%%%%%%%%%%%%%%%%%%%%%%%%%%%%%%%%%%%%%%%%%%%%%%%%%%%%%%%%%%%%%%%%%%%%%%%%%%%%%%%%%%%%

\BLAU{We now find the Lie algebra $\mathfrak{meta}(1,d)$ in more than one spatial dimension $d$. 
We begin with} generic conditions which will hold for any dimension $d>1$. 
Specific results apply for $d=2$ and will be presented in section~4.

We start from the ansatz (\ref{X1ddim}), with $p\ne 1$.
Throughout, we shall assume $\vec{B}=\vec{0}$, unless explicitly stated otherwise.
{}From the defining commutator relation, we have the generator
\BEA
Y_0^j & := & \demi[X_1,Y_{-1}^j]\nonumber\\
      &  = & -\alpha r_j\partial_t-\left(t+\demi(1-p)(\vec{\beta}\cdot\vec{r})\right)\partial_{r_j}-pr_j\vec{\beta}\cdot
             \partial_{\vec{r}} -\demi(1-p)\beta_j\vec{r}\cdot\partial_{\vec{r}}-(k/2)\gamma_j ~~~~
\label{y0j}
\EEA
To be specific, let $d=3$, but the conclusions will apply to any $d\geq 2$.
Take from (\ref{detalpha}) the value $\alpha=\frac{1}{4}(p-1)(p+1)\vec{\beta}^2$ and work out $[Y_0^j,Y_0^i]$ for $i\ne j$.
For example
\BEA
     [Y_0^x,Y_0^y] &  = & \frac{(3p-1)(p+1)(p-1)}{8}(\beta_yx-\beta_xy)\partial_t\nonumber\\
                   &    & +\left(\frac{(3p-1)(p+1)}{4}(\beta_x\beta_yx-\beta_x^2y)+\frac{(1-p)^2}{4}(\beta_z^2y-\beta_y\beta_zz)\right)
                          \partial_x\nonumber\\
                   &    & +\left(\frac{(3p-1)(p+1)}{4}(\beta_y^2x-\beta_x\beta_yy)-\frac{(1-p)^2}{4}(\beta_z^2x-\beta_x\beta_zz)\right)
                          \partial_y\nonumber\\
                   &    & +p^2(\beta_y\beta_zx-\beta_x\beta_zy)\partial_z \label{1newcom}
\EEA
which must be expressed in terms of the generators of the Lie algebra under construction, including 
the rotation generators $R_{ij}=r_i\partial_{r_j}-r_j\partial_{r_i}$.\footnote{The {\em level} $\mathpzc{x}$ of a generator
$\cal X$ is defined by $[X_0, {\cal X}]=\mathpzc{x} {\cal X}$. Hence the commutator of the level-zero generators $Y_0^{xy}$ must
itself be of level zero, hence be a linear combination of $X_0$, $Y_0^j$ or $R_{ij}$.}  
\BLAU{From (\ref{1newcom}) we see that the
commutator $[Y_0^x,Y_0^y]$ only becomes a linear combination of known generators if the parameter $p$ obeys} %the equation
\BEQ \label{haracteristiceq}
p^2-\frac{1}{4}(1-p)^2 = \frac{1}{4}(p+1)(3p-1)\stackrel{!}{=}0,
\EEQ

\noindent
\BLAU{{\bf Proposition 2:} {\em Consistent representations of $\mathfrak{meta}(1,d)$ with $d>1$ 
are only possible in the cases (i) $p_1=-1$ and (ii) $p_2=\frac{1}{3}$.}}

\BLAU{In either of these cases,\footnote{In contrast with $1D$ meta-conformal transformations 
(see sect.~2), they are obtained here without any normalisation condition.} the commutator (\ref{1newcom}) simplifies to} 
\begin{subequations}
\begin{align}
[Y_0^x,Y_0^y] &= -p^2\left(\beta_z^2 R_{xy}+\beta_x\beta_z R_{yz}+\beta_y\beta_z R_{zx}\right) \label{y0xy}
\end{align}
Similarly, for the same values of $p=-1,\frac{1}{3}$, we find (still for $d=3$)
\begin{align}
     & [Y_0^y,Y_0^z] = -p^2\left(\beta_x\beta_z R_{xy}+\beta_x^2 R_{yz}+\beta_x\beta_y R_{zx}\right) \label{y0yz}\\
     & [Y_0^z,Y_0^x] = -p^2\left(\beta_y\beta_z R_{xy}+\beta_x\beta_y R_{yz}+\beta_y^2 R_{zx}\right) \label{y0xz}
\end{align}
\end{subequations}
\BLAU{Therefore, a discussion of rotation-invariance is necessary.}
\begin{enumerate}
\item One might choose to keep full spatial rotation-invariance, with all three generators $R_{xy}$, $R_{yz}$, $R_{zx}$.
Since the invariant equation (\ref{Boltzmannd}) contains a vector proportional
to $\vec{\beta}$, one must include into the rotation generators, viz. $R_{ij}\mapsto \bar{R}_{ij}$,
terms which describe the simultaneous rotations of the position
$\vec{r}$ and of $\vec{\beta}$. However, changing $\vec{\beta}$ then implies changing the invariant equation.
The transformations found will map one equation of the type (\ref{Boltzmannd}) to another equation of the same type.
\item Here, we shall use rotation-invariance to orient the coordinate axes such that $\vec{\beta}$
is along the $x$-axis. In other words,
we shall fix, from now on, $\beta_x=\beta\ne 0$ and $\beta_y=\beta_z=\ldots=0$.
Explicit rotation-invariance will only apply to rotations which leave the
$x$-axis invariant. These do not exist for $d=2$, but for $d=3$ we have the rotation \BLAU{$R=R_{yz}$}.
\end{enumerate}

\subsection{Meta-conformal algebra in $d=3$ dimensions with $\vec{\gamma}=\vec{0}$}

The case of $d=3$ spatial dimensions gives the generic structure meta-conformal transformations.
Throughout, we shall fix $\vec{\beta}=(\beta,0,0)$. First, we restrict to the more simple case $\vec{\gamma}=\vec{0}$.
The rotation generator is $R_{ij}=r_i\partial_{r_j}-r_j\partial_{r_i}$.
With (\ref{detalpha}), we have %%$\alpha = \frac{1}{4}{(p+1)(p-1)}\beta^2$. Explicitly
\BEA
X_1 & = & -\left(t^2+\alpha(x^2+y^2+z^2)\right)\partial_t-\left(2tx+\beta x^2+\beta p(y^2+z^2)\right)\partial_x\nonumber\\
    &   & -(2t+\frac{1-p}{2}\beta x)y\partial_y-(2t+\frac{1-p}{2}\beta x)z\partial_z-2\delta t.       \label{X1d3A1}
\EEA
All other generators can be found from (\ref{X1d3A1}). Starting from $Y_0^j:=\demi[X_1,Y_{-1}^j]$, 
we check\footnote{For further calculational details, 
see the preprint version {\tt arXiv:1810.09855v2}, or \cite{Stoimenov19}.\label{ftn:calc}} 
that $[Y_0^x, Y_0^y]=[Y_0^x, Y_0^z]=0$.
In addition, if we take either $p=-1$ or $p=\frac{1}{3}$, then
\BEQ
[Y_0^y, Y_0^z] = -p^2\beta^2R_{yz}\label{1ntrivcom0}
\EEQ
does not vanish for $d\geq 3$, see (\ref{y0yz}). Also, $[X_1, R_{yz}]=0$, as expected from rotation-invariance. The next family
of generators is obtained from $Y_1^j :=[X_1,Y_{0}^j]$.  
If $n, m\in \{-1,0,1\}$ and $j=x,y,z$, the non-vanishing commutators are compactly written as
\BEA && [X_n, X_m]=(n-m)X_{n+m} \;\; , \quad [X_n,Y_m^j]=(n-m)Y_{n+m}^j\nonumber\\
     && [Y_n^x, Y_m^x]=(n-m)(\alpha X_{n+m}+\beta Y_{n+m}^x)\nonumber\\
     && [Y_n^y, Y_m^y]=[Y_n^z, Y_m^z]=(n-m)(\alpha X_{n+m}+p\beta Y_{n+m}^x)\nonumber\\
     && [Y_n^x, Y_m^{w}]=[Y_n^{w}, Y_m^x]=(n-m)\frac{1-p}{2}\beta Y_{n+m}^{w} ~~\mbox{\rm ;~~ for $w=y,z$,}\nonumber\\
     && [Y_n^y, Y_m^z]=\delta_{n+m,0}(n-m-\delta_{nm})p^2\beta^2R_{yz},\label{pcommcomf13}\\
%     && [Y_n^z, Y_m^y]=\delta_{n+m,0}(n-m+\delta_{nm})p^2\beta^2R_{yz},\nonumber\\
     && [Y_m^y, R_{yz}]=Y_m^z \;\; , \quad [Y_m^z, R_{yz}]= -Y_m^y.\nonumber
\EEA

\noindent {\bf Proposition 3:}
{\em If $p=-1$ or $p=\frac{1}{3}$, and $\vec{\gamma}=\vec{0}$, the set 
$\mathfrak{meta}(1,3) := \langle X_{0,\pm 1}, Y_{0,\pm 1}^{x,y,z}, R_{yz}\rangle$
of differential operators, as derived from (\ref{X1d3A1}), closes into a meta-conformal Lie algebra,
whose structure is determined by the commutators (\ref{pcommcomf13}),
with two distinct representations.}

\noindent{\bf Proof:} The first part of the proposition follows from the closure of the commutators (\ref{pcommcomf13}).
For the  second part, let $p=\frac{1}{3}$ and consider the commutators (\ref{pcommcomf13}). 
Re-define the generators $Y_n^x\mapsto {\cal Y}_n^x := -\frac{2}{3}\beta X_n+Y_n^x$. Again, the commutators (\ref{pcommcomf13})
will hold, where $Y_n^x$ is substituted by ${\cal Y}_n^x$ and if one replaces therein $\beta \mapsto -(\beta/3)$. \hfill q.e.d.

\BLAU{For clarity, we shall mainly concentrate on the most simple representation, with $p=-1$. For easy reference, we repeat below 
the commutators of the $p=-1$ representation of $\mathfrak{meta}(1,3)$} (here  $n, m\in \{-1,0,1\}$ and $j=x,y,z$)
\BEA && [X_n, X_m]=(n-m)X_{n+m} \;\; , \quad [X_n,Y_m^j]=(n-m)Y_{n+m}^j \label{sumdirect} \nonumber \\
     && [Y_n^x, Y_m^x]=(n-m)\beta Y_{n+m}^x\nonumber\\
     && [Y_n^y, Y_m^y]=[Y_n^z, Y_m^z]=-(n-m)\beta Y_{n+m}^x\nonumber\\
     && [Y_n^x, Y_m^{w}]=[Y_n^{w}, Y_m^x]=(n-m)\beta Y_{n+m}^{w} ~~\mbox{\rm ;~~ for $w=y,z$,}\nonumber\\
     && [Y_n^y, Y_m^z]=\delta_{n+m,0}(n-m-\delta_{nm})\beta^2R_{yz}, \label{defcommcomf13}\\
%     && [Y_n^z, Y_m^y]=\delta_{n+m,0}(m-n+\delta_{nm})\beta^2R_{yz},\nonumber\\
     && [Y_m^y, R_{yz}]=Y_m^z \;\; , \quad [Y_m^z, R_{yz}]= -Y_m^y.\nonumber
\EEA

\BLAU{The structure of meta-conformal algebras can be further simplified. This is in contrast with the semi-direct sums which hold
true e.g. for the Schr\"odinger algebra.} 

\noindent\BLAU{{\bf Proposition 4:}
{\em The Lie algebra $\mathfrak{meta}(1,3)$ decomposes as a direct sum}
\BEQ 
\mathfrak{meta}(1,3) %\cong \mathfrak{sl}(2,\mathbb{R})\oplus\mathfrak{g} 
\cong \mathfrak{sl}(2,\mathbb{R})\oplus\mathfrak{conf}(3) \cong \mathfrak{sl}(2,\mathbb{R})\oplus B_2.\label{structure13}
\EEQ}

\begin{figure}[tb]
\begin{center}
\includegraphics[width=.5\hsize]{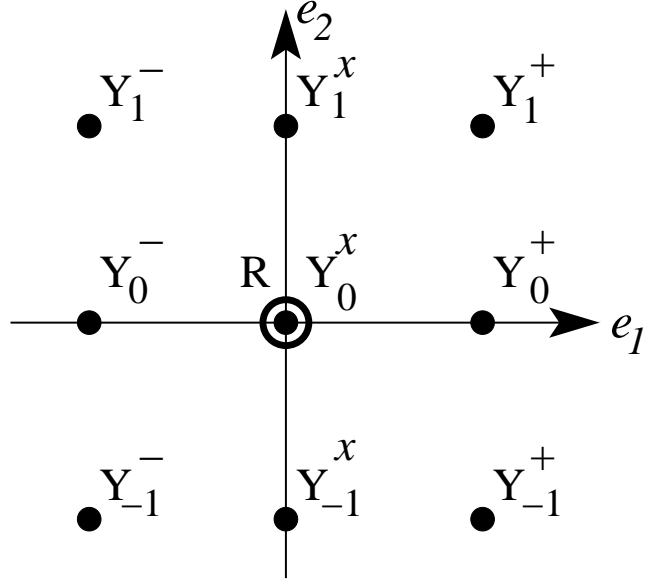}
\end{center}
\caption{\small Root diagramme of the complex Lie algebra $B_2$ and the correspondence with generators of $3D$ 
meta-conformal transformations.   \label{figracines}}
\end{figure}

\noindent\BLAU{{\bf Proof:} The relations (\ref{defcommcomf13}) %(\ref{sumdirect}) 
imply that $\langle X_{0,\pm 1}\rangle\cong\mathfrak{sl}(2,\mathbb{R})$. In addition, the action of
the $\mathfrak{sl}(2,\mathbb{R})$-subalgebra on the generators of the subalgebra 
$\mathfrak{g} := \langle Y_m^{x,y,z}, R\rangle_{m\in\{\pm 1,0\}}$, with $R=R_{yz}$ shows a semi-direct structure.  
Changing the base of the Lie algebra according to $X_n\mapsto A_n := X_n - Y_n^x/\beta $, one has 
$\langle A_{\pm 1,0}\rangle\cong \mathfrak{sl}(2,\mathbb{R})$ and the direct sum
$\mathfrak{meta}(1,3)\cong\mathfrak{sl}(2,\mathbb{R}) \oplus \mathfrak{g}$, as can be checked through the commutators 
\BD 
[A_n, Y_m^x]=[A_n, Y_m^y]=[A_n, Y_m^z]=[A_n, R]=0.\nonumber
\ED
The structure of the Lie sub-algebra $\mathfrak{g}$ is made clear by defining the generators
\BEQ 
Y_n^+=Y_n^y+\II Y_n^z \;\; , \;\; Y_n^-=Y_n^y-\II Y_n^z.\label{defplusminus} \nonumber
\EEQ
Then the non-vanishing commutators of the Lie algebra $\mathfrak{g}$ become
\BD 
[Y_n^x, Y_m^x]=(n-m)\beta Y_{n+m}^x \;\; , \;\; [Y_n^x, Y_m^{\pm}]=(n-m)\beta Y_{n+m}^{\pm} 
                                   \;\; , \;\; [R, Y_m^{\pm}]=\pm \II Y_m^{\pm}
\ED
\BEQ
[Y_n^+, Y_m^-] = \left\{
\begin{array}{lll} 2(n-m)\beta Y_{n+m}^x                     & \mbox{\rm ~~;~ if $n+m\ne 0$} \\
                   2\II\beta^2 R                             & \mbox{\rm ~~;~ if $n=m=0$} \\
                   2(n-m)\beta Y_{n+m}^x-4\II \beta^2 R      & \mbox{\rm ~~;~ if $n\ne m$} \label{minusplus} \nonumber 
\end{array} \right.
\EEQ
The correspondence with the roots of the complex Lie algebra $B_2$ is shown in figure~\ref{figracines}.\hfill q.e.d.} 

%%====================================================================================================================================%%
%%===================================================================================================================================%%

\BLAU{Although we did not carry out the explicit construction of the generators for $d>3$ dimensions, 
counting their number allows to formulate the following}

\noindent\BLAU{{\bf Conjecture:} {\em In $d\geq 1$ spatial dimensions, one has for $\mathfrak{meta}(1,d)$ 
the Lie algebra isomorphisms}
\BEQ \label{gl:confmeta}
\mathfrak{meta}(1,1) \cong A_1 \oplus A_1 \;\; , \;\;
\mathfrak{meta}(1,d) \cong \mathfrak{sl}(2,\mathbb{R}) \oplus \mathfrak{conf}(d) \cong 
\left\{ \begin{array}{ll} A_1 \oplus D_{n+1} & \mbox{\rm ~~;~ if $d=2n$} \\ 
                          A_1 \oplus B_{n+1} & \mbox{\rm ~~;~ if $d=2n+1$} 
        \end{array} \right.
\EEQ
%%MH vieille version, je ne le trouve pas aussi informative que la nouvelle 
%\BEQ
%\mathfrak{meta}(1,1) \cong A_1 \oplus A_1 \;\; , \;\;
%\mathfrak{meta}(1,2n) \cong A_1 \oplus D_{n+1} \;\; , \;\;
%\mathfrak{meta}(1,2n+1) \cong A_1 \oplus B_{n+1} 
%\EEQ
{\em with $n=1,2,\ldots$, and where $\mathfrak{conf}(d)$ is the ortho-conformal Lie algebra in $d$ dimensions and 
$A_1$, $B_n$, $D_n$ are simple complex Lie algebras from Cartan's classification (and $D_2\cong A_1\oplus A_1$).}}

\BLAU{In stating this, we anticipate results on $\mathfrak{meta}(1,2)$ from sect.~4.
}

A further important difference between $d=2$ and $d\geq 3$ dimensions arises from the non-vanishing of commutators
such as (\ref{1ntrivcom0}) when $d\geq 3$. \BLAU{The vanishing of these commutators for $d=2$ is the reason why
an infinite-dimensional extension of the $2D$ meta-conformal Lie algebra can be constructed,}
as we shall show in section~4.

\subsection{Meta-conformal algebra in $d=3$ dimensions with $\vec{\gamma}\ne\vec{0}$}

We first redefine the generator of rotations, which now should include rotations of $\vec{\gamma}$
\BEQ
R_{yz}\mapsto \bar{R}_{yz} =: y\partial_z-z\partial_y+\gamma_y\partial_{\gamma_z}-\gamma_z\partial_{\gamma_z}\label{gRyz}.
\EEQ
Here, we must also include the term $\vec{B}\ne \vec{0}$.
To do so, we modify $X_1$ of eq.~(\ref{X1d3A1}) by the following ansatz, according to (\ref{X1ddim})
\BEA X_1 & \mapsto &  X_1+\tilde{X}_1
\nonumber\\
     \tilde{X}_1 & = & -\left( a(\vec{\beta}\cdot\vec{r})\vec{\gamma}
     +b(\vec{\gamma}\cdot\vec{r})\vec{\beta}
     +c(\vec{\beta}\cdot\vec{\gamma})\vec{r}\right)\cdot\partial_{\vec{\gamma}}-k(\vec{\gamma}\cdot\vec{r})
\nonumber\\
                 & = & -\beta\left((a+b+c)x\gamma_x+a(y\gamma_y+z\gamma_z)\right)\partial_{\gamma_x}-k(x\gamma_x+y\gamma_y+z\gamma_z)
\nonumber\\
                 &   & -\beta(bx\gamma_y+cy\gamma_x)\partial_{\gamma_y}-\beta(bx\gamma_z+cz\gamma_x)\partial_{\gamma_z}\label{modX1A1}
\EEA
where $a,b,c$ and $k$ are constants to be determined.
Next, we construct $Y_0^{x,y,z}$ and $Y_1^{x,y,z}$, as usual.\fnref{ftn:calc} We find the explicit extra terms beyond (\ref{y0j})
\begin{subequations} \label{modY0-A1}
\begin{align}
     & Y_0^x \mapsto Y_0^x+\tilde{Y}_0^x, \quad Y_0^y \mapsto Y_0^y+\tilde{Y}_0^y, \quad Y_0^z \mapsto Y_0^z+\tilde{Y}_0^z\nonumber\\
     & \tilde{Y}_0^x = -(\beta/2)\left((a+b+c)\gamma_x\partial_{\gamma_x}+b(\gamma_y\partial_{\gamma_y}
                        +\gamma_z\partial_{\gamma_z}\right)-(k/2)\gamma_x\label{modY0xA1}\\
     & \tilde{Y}_0^y = -(\beta/2)(a\gamma_y\partial_{\gamma_x}+c\gamma_x\partial_{\gamma_y})-(k/2)\gamma_y\label{modY0yA1}\\
     & \tilde{Y}_0^z = -(\beta/2)(a\gamma_z\partial_{\gamma_x}+c\gamma_x\partial_{\gamma_z})-(k/2)\gamma_z\label{modY0zA1}
\end{align}
\end{subequations}
The values of the constants $a,b,c$ are fixed from the requirement that the Lie algebra commutators of 
$\mathfrak{meta}(1,3)$ are those of the case $\vec{\gamma}=\vec{0}$ treated
above.\footnote{Clearly, modifying $X_1$ and correspondingly $Y_0^{x,y,z}$ and $Y_1^{x,y,z}$
by {\em additive} terms does not change the commutation relations.} Eqs.~(\ref{modY0-A1}) imply
\begin{enumerate}
\item the conditions $[Y_0^x,Y_0^y]=[\tilde{Y}_0^x,\tilde{Y}_0^y]=0$ yield $a=b=-c$.
\item the condition $[Y_0^y,Y_0^z]=-p^2\beta^2\bar{R}_{yz}$ yields $a=\pm 2p$.
\end{enumerate}

\noindent
\BLAU{{\bf Proposition 5:} {\em The time-space $3D$ meta-conformal transformations with $\vec{\gamma}\ne\vec{0}$ 
can be labelled by the pair $(p,a)$. There are 
four possibilities: $(p,a)=(-1,-2)$, $(-1,2)$, $(\frac{1}{3},\frac{2}{3})$, $(\frac{1}{3},-\frac{2}{3})$. 
The value of the constant $k$ is not fixed by the commutators.}}

The commutator $Y_1^j = [X_1, Y_0^j]$ gives the corresponding extensions of the generators $Y_1^j$. 

\subsection{Symmetries of the ballistic transport equation}

\BLAU{Returning to the linear ballistic transport equation, in the form (\ref{Boltzmannd},\ref{eqsym})}
\BEQ \label{inveqA}
\hat{\mathscr{S}}\vph_{\vec{\gamma}}(t,\vec{r})=\left(\partial_t+\frac{2}{\beta(p-1)}\partial_x\right)\vph_{\vec{\gamma}}(t,\vec{r})=0.
\EEQ
\BLAU{we can state the conditions for it having a meta-conformal dynamical symmetry, 
for $\vec{\gamma}\ne\vec{0}$ and $\vec{B}\ne \vec{0}$.} 
Herein, the solution $\vph=\vph_{\vec{\gamma}}(t,\vec{r})$ may also depend on the vector $\vec{\gamma}$ of `rapidities'.

\noindent {\bf Proposition 6:} {\em For generic dimension $d>2$, and $p=\frac{1}{3}$ or $p=-1$, we have
\begin{itemize}
\item[(i)] For $\vec{\gamma}=\vec{0}$, the meta-conformal representations constructed above 
leave invariant the solution space of the equations (\ref{inveqA}), under the condition $\delta=0$.
\item[(ii)] For $\vec{\gamma}\ne \vec{0}$ and if $\vph_{\vec{\gamma}}(t,\vec{r})=\vph(t,\vec{r})$
does not explicitly depend on $\vec{\gamma}$,
the corresponding meta-conformal representation leaves the solution space of (\ref{inveqA}) invariant,
if $k=1$ and $\gamma_x=(1-p)\beta\delta$.
\item[(iii)] If the solution $\vph_{\vec{\gamma}}(t,\vec{r})$ does also depend on $\vec{\gamma}$,
invariance of the solution space of (\ref{inveqA}) is only obtained under the conditions $k=1$ and
\BEQ
\left(\delta +\frac{1}{\beta(p-1)}\gamma_x +
\frac{a}{p-1}\vec{\gamma}\cdot\partial_{\vec{\gamma}}\right)\vph_{\vec{\gamma}}(t,\vec{r})=0\label{condsym}
\EEQ
\end{itemize}
}
\noindent In case (iii) we have an on-shell or a {\em conditional symmetry} of eq.~(\ref{inveqA}),
see e.g. \cite{Bluman69,Fushchych93,Cherniha17}.

\noindent
{\bf Proof:} \BLAU{This follows from the non-vanishing commutators} 
\BEA
\left[\hat{\mathscr{S}},X_0\right]   &=& -\hat{\mathscr{S}} \;\; , \hspace{2.5truecm}\;\; \left[\hat{\mathscr{S}},Y_0^x\right]
                                         =-\frac{p+1}{2}\beta \hat{\mathscr{S}}
\nonumber \label{cond0sym}\\
\left[\hat{\mathscr{S}},X_1\right]   &=& -\left(2t+(p+1)\beta x\right)\hat{\mathscr{S}}-2\left(\delta +\frac{k}{\beta(p-1)}\gamma_x
                                   +\frac{a}{p-1}\vec{\gamma}\cdot\partial_{\vec{\gamma}}\right)
\nonumber \label{cond1sym}\\
\left[\hat{\mathscr{S}},Y_1^x\right] &=& -\frac{\beta(p+1)}{2}\left(2t+(p+1)\beta x\right)\hat{\mathscr{S}} \nonumber\\
                              & & -(p+1)\beta\left(\delta+\frac{2+(p-1)k}{(p-1)(p+1)\beta}\gamma_x
                                  + \frac{a}{p-1}\vec{\gamma}\cdot\partial_{\vec{\gamma}}\right)
\nonumber \label{cond2sym}
\EEA
and recalling that $a=\pm 2p$. \hfill q.e.d.

%%%%%%%%%%%%%%%%%%%%%%%%%%%%%%%%%%%%%%%%%%%%%%%%%%%%%%%%%%%%%%%%%%%%%%%%%%%%%%%%%%%%%%%%%%%%%%%%%%%%%%%%%%%%%%%%%%%%%%%%%%%%%%%%%%%%%%%%
\section{Meta-conformal algebra in $d=2$ spatial dimensions}
%%%%%%%%%%%%%%%%%%%%%%%%%%%%%%%%%%%%%%%%%%%%%%%%%%%%%%%%%%%%%%%%%%%%%%%%%%%%%%%%%%%%%%%%%%%%%%%%%%%%%%%%%%%%%%%%%%%%%%%%%%%%%%%%%%%%%%%%

\BLAU{Finding meta-conformal transformations} in $d=2$ space dimensions (with points $(t,x,y)\in\mathbb{R}^3$) proceeds as follows.
As in sections~2 and~3, the generators of translations and dilatations read
\begin{subequations} \label{gensimple}
\begin{align}
X_{-1}   &= -\partial_t                                     \label{timetranslations2}\\
Y^x_{-1} &= -\partial_x \;\; , \;\;  Y^y_{-1} = -\partial_y \label{xyspacetranslations}\\
X_0      &= -t\partial_t-x\partial_x-y\partial_y-\delta.    \label{xydynscaling}
\end{align}
\end{subequations}
The form of $X_1$ is given by (\ref{X1ddim}) where for simplicity we set $\vec{B}=\vec{0}$ and fixed $k=2$.
%We shall show that a closed and infinite-dimensional Lie algebra of dynamical symmetries can be found.
Explicitly
\BEA
     X_1 & = & -\left(t^2+\alpha(x^2+y^2)\right)\partial_t-\left(2tx+\beta_xx^2+(1-p)\beta_yxy+p\beta_xy^2\right)\partial_x\nonumber\\
         &   & -\left(2ty+p\beta_yx^2+(1-p)\beta_xxy+\beta_yy^2\right)\partial_y-2\delta t -2\gamma_xx-2\gamma_yy.\label{xyX1}
\EEA
and \BLAU{we kept the vector $\vec{\beta}=(\beta_x,\beta_y)$ arbitrary. 
In complete analogy with the case of $d\geq 3$ dimensions, 
the generators of the nine-dimensional Lie algebra 
$\mathfrak{meta}(1,2)=\left\langle X_n, Y_n^j\right\rangle_{n\in\{\pm 1,0\},j\in\{x,y\}}$ are found,\fnref{ftn:calc} 
for the two admissible cases (i) $p=-1$ and (ii) $p=\frac{1}{3}$. 
However, the Lie algebra of $2D$ meta-conformal transformations is infinite-dimensional, as we shall now show.} 

\subsection{Infinite-dimensional extension}
%
%%++++++++++++++++++++++++++++++++++++++++++++++++++++++++++++++++++++++++++++++++++++++++++++++++++++++++++++++++++++++++++++++++++++%%
\begin{table}
\begin{center}\begin{tabular}{|ll|lll|} \hline
transformation &            & \multicolumn{1}{c}{$\tau$} & \multicolumn{1}{c}{$w$} & \multicolumn{1}{c|}{$\bar{w}$}        \\ \hline
meta-conformal $2D$  \hfill {\sc i}
               & $\alpha=0$    & $t$                     & $t+\beta (x+\II y)$     & $t+\beta (x-\II y)$                 \\[0.15truecm]
meta-conformal $2D$  \hfill{\sc ii}
               & $\alpha\ne 0$ & $t+\frac{2\beta}{3}\,x$ & $t+\frac{\beta}{3}\,(x-\II y)$ & $t+\frac{\beta}{3}\,(x+\II y)$ \\ \hline
\end{tabular}\end{center}
\caption[tab3]{\small Possible choices for the `time' and `complex' light-cone coordinates $\tau,w,\bar{w}$ of the
$2D$ meta-conformal generators (\ref{gl4.3}) in $d=2$ spatial dimensions,
in terms of the time-space coordinates $t,x,y$.  \label{tab3}}
\end{table}
%%++++++++++++++++++++++++++++++++++++++++++++++++++++++++++++++++++++++++++++++++++++++++++++++++++++++++++++++++++++++++++++++++++++%%
\BLAU{In what follows, we shall choose coordinate axes such that $\vec{\beta}=(\beta,0)$.}

\noindent
\BLAU{
{\bf Proposition 7:} {\em Consider the set of generators, with $n\in\mathbb{Z}$}
\BEA
A_n   &=& - \tau^{n+1}\partial_{\tau} - (n+1) \vartheta\, \tau^n \nonumber \\
B_n^+ &=& - w^{n+1}\partial_{w} -(n+1) \Delta\, w^n \label{gl4.3} \\
B_n^- &=& - \bar{w}^{n+1}\partial_{\bar{w}} - (n+1) \overline{\Delta}\, \bar{w}^n \nonumber 
\EEA
{\em which act on the time-space coordinates $(\tau,w,\bar{w})$. Their non-vanishing commutators are}
\BEQ \label{gl4.4}
\left[ A_n, A_m \right] = (n-m) A_{n+m} \;\; , \;\;
\left[ B_n^{\pm} , B_m^{\pm} \right] = (n-m) B_{n+m}^{\pm}
\EEQ
{\em The two possible cases of $2D$ meta-conformal transformations correspond to (i) $p=-1$ with $\alpha=0$ and 
(ii) $p=\frac{1}{3}$ with $\alpha = -\frac{2}{9}\vec{\beta}^2\ne 0$. The relationship with the usual time-space coordinates 
$(t,x,y)$ is given, for both cases, in table~\ref{tab3} and the meta-conformal weights are in table~\ref{tab4}.}
}

\noindent
\BLAU{{\bf Proof:} straightforward calculation. The correspondence with the representations of the finite-dimensional
sub-algebra $\mathfrak{meta}(1,2)$ will be given in eqs.~(\ref{glcas1corr},\ref{glcas2corr}) below.  \hfill q.e.d.} 

\BLAU{The Lie algebra (\ref{gl4.4}) is isomorphic to the direct sum 
$\mathfrak{vect}(S^1)\oplus\mathfrak{vect}(S^1)\oplus\mathfrak{vect}(S^1)$, the direct
sum of three centre-less Virasoro algebras. The corresponding {\em meta-conformal weights} $\vartheta$, $\Delta$, 
$\overline{\Delta}$ are expressed in table~\ref{tab4} in terms of the scaling dimension $\delta$ and the two `rapidities'
$\gamma_{x,y}$.} 

%%++++++++++++++++++++++++++++++++++++++++++++++++++++++++++++++++++++++++++++++++++++++++++++++++++++++++++++++++++++++++++++++++++++%%
\begin{table}
\begin{center}\begin{tabular}{|ll|lll|} \hline
transformation &  & \multicolumn{1}{c}{$\vartheta$} & \multicolumn{1}{c}{$\Delta$} 
                  & \multicolumn{1}{c|}{$\overline{\Delta}$}        \\ \hline
meta-conformal $2D$  \hfill {\sc i}
               & $\alpha=0$    & $\delta-\gamma/\beta$                                      & $\gamma/\beta$     
               & $\bar{\gamma}/\beta$                   \\[0.15truecm]
meta-conformal $2D$  \hfill{\sc ii}
               & $\alpha\ne 0$ & $\frac{3\gamma}{\beta}+\frac{3\bar{\gamma}}{\beta}-\delta$ & $\delta-\frac{3\gamma}{\beta}$ 
               & $\delta-\frac{3\bar{\gamma}}{\beta}$  \\ \hline
\end{tabular}\end{center}
\caption[tab3]{\small Possible choices for the `meta-conformnal weights' $\vartheta$, $\Delta$, $\overline{\Delta}$ for the
$2D$ meta-conformal generators (\ref{gl4.3}), in terms of the scaling dimension $\delta$ and the rapidities $\gamma_x,\gamma_y$,
where $\gamma=\demi(\gamma_x + \II\gamma_y)$ and $\bar{\gamma}=\demi(\gamma_x - \II\gamma_y)$.   \label{tab4}}
\end{table}
%%++++++++++++++++++++++++++++++++++++++++++++++++++++++++++++++++++++++++++++++++++++++++++++++++++++++++++++++++++++++++++++++++++++%%

\BLAU{The finite transformations associated with the generators $A_n,B_n^{+},B_n^{-}$ with $n\in\mathbb{Z}$
are given by the corresponding Lie series.
With the definition $\vph(\tau,w,\bar{w})=\phi(t,z,\bar{z})$, the final result is
\begin{subequations} \label{finit-2D}
\begin{align}
\hspace{-0.3truecm}B_n^+:&\quad \vph'(\tau,w,\bar{w}) = \left(\frac{\D w'}{\D w}\right)^{\Delta}\vph(\tau',w',\bar{w}') \;\; ;\;\;
                              \tau'=\tau\;\;\hspace{0.38truecm},\;\; w'=f(w)\;\;,\;\; \bar{w}'=\bar{w} \\
\hspace{-0.3truecm}B_n^-:&\quad \vph'(\tau,w,\bar{w}) 
= \left(\frac{\D \bar{w}'}{\D \bar{w}}\right)^{\overline{\Delta}}\vph(\tau,w,\bar{w}') 
        \;\;\hspace{0.2truecm};\;\;
        \tau'=\tau\;\;\hspace{0.38truecm},\;\; w'  =  w\;\;\hspace{0.45truecm},\;\;  \bar{w}' = \bar{f}(\bar{w})\\
\hspace{-0.3truecm}A_n:&\quad \vph'(\tau,w,\bar{w})  
=  \left(\frac{\D \tau'}{\D \tau}\right)^{\vartheta}\vph\left(\tau',w',\bar{w}'\right) \;\;;\;\;
\tau' = b(\tau)\;\;,\;\;  w'=w \;\;\hspace{0.38truecm}, \;\;  \bar{w'}=\bar{w}.
\end{align}
\end{subequations}
where $f=f(w)$, $\bar{f}=\bar{f}(\bar{w})$ and $b=b(\tau)$ are arbitrary differentiable functions. The  differential
equations for the Lie series are recovered by expanding $b(t)=t-\eps t^{n+1}$, and analogously for $f(z)$ and $\bar{f}(\bar{z})$.} 

\BLAU{Eqs.~(\ref{finit-2D}) show that the relaxational behaviour described by the
$2D$ meta-conformal symmetry is governed by {\em three} independent
conformal transformations, rather than two as it is the case for $2D$ ortho-conformal invariance at equilibrium.}

\noindent
\BLAU{
{\bf Proposition 8:} {\em For a vanishing meta-conformal weight $\vartheta=\vartheta_{\phi}=0$, 
the Lie algebra acts as a dynamical symmetry on the linear ballistic transport equation $\hat{\mathscr{S}}\phi(t,z,\bar{z})=0$ with}
\BEA
\hat{\mathscr{S}} = \left\{\begin{array}{ll}
-\partial_t + \frac{1}{\beta} \bigl( \partial_z + \partial_{\bar{z}} \bigr) & \mbox{\rm ~~;~ case {\sc i} with $p=-1$} \\[0.15truecm]
\partial_t - \frac{3}{\beta} \bigl( \partial_z + \partial_{\bar{z}} \bigr) & \mbox{\rm ~~;~ case {\sc ii} with $p=\frac{1}{3}$}
\end{array} \right.
\EEA
\noindent
{\bf Proof:} this follows from 
$[ A_n , \hat{\mathscr{S}} ] = (n+1)\tau^n \hat{\mathscr{S}} - (n+1)n \vartheta\, \tau^n$ and 
$[ B_n^{\pm}, \hat{\mathscr{S}} ] = 0$. \hfill q.e.d. 
}

\subsection{The case $p=-1$}
\subsubsection{Lie algebra generators}

\BLAU{The tables~\ref{tab3} and~\ref{tab4} give the relationship between the coordinates $(\tau,w,\bar{w})$ and the usual
time-space coordinates $(t,x,y)$ and also the explicit meta-conformal weights. In this sub-section, we consider the case $\alpha=0$. 
The correspondence between the generators is as follows
\BEQ \label{glcas1corr}
A_n = X_n - \frac{1}{\beta} Y_n^x \;\; , \;\; B_n^{\pm} = \frac{1}{2\beta} \left( Y_n^x \pm \II Y_n^y\right)
\EEQ
For illustration, we write down explicitly the generators of $\mathfrak{meta}(1,2)$ 
in the original, physically motivated, basis. In addition, 
although the discussion above was done for the choice $\vec{\beta}=(\beta,0)$, we give here the generators for the generic
situation where $\vec{\beta}=(\beta_x,\beta_y)$. They read\fnref{ftn:calc}} 
\begin{subequations} \label{genaddi}
\begin{align}
  X_1   &= -t^2\partial_t-\left(2tx+\beta_xx^2+2\beta_yxy-\beta_xy^2\right)\partial_x\nonumber\\
        &  ~~~-\left(2ty-\beta_yx^2+2\beta_xxy+\beta_yy^2\right)\partial_y-2\delta t -2\gamma_xx-2\gamma_yy\label{rightX1}\\
  Y_0^x &=  -(t+\beta_xx+\beta_yy)\partial_x-(\beta_xy-\beta_yx)\partial_y-\gamma_x\label{righty0x}\\
  Y_0^y &= -(\beta_yx-\beta_xy)\partial_x-(t+\beta_yy+\beta_xx)\partial_y-\gamma_y\label{righty0y}\\
Y^x_1 &= -\left(t^2+2t\beta_xx+2t\beta_yy+(\beta_x^2-\beta_y^2)x^2+4\beta_x\beta_yxy-(\beta_x^2-\beta_y^2)y^2\right)\partial_x
\nonumber\\
&  ~~~-\left(2t\beta_xy-2t\beta_yx-2\beta_x\beta_yx^2+2(\beta_x^2-\beta_y^2)xy+2\beta_x\beta_yy^2\right)\nonumber\\
&  ~~~-2\gamma_x(t+\beta_xx+\beta_yy)-2\gamma_y(\beta_xy-\beta_yx)\label{righty1x}\\
      Y^x_1 &=
      -\left(2t\beta_yx-2t\beta_xy+2\beta_x\beta_yx^2-2(\beta_x^2-\beta_y^2)xy-2\beta_x\beta_yy^2\right)\partial_x\nonumber\\
      &  ~~~-\left(t^2+2t\beta_xx+2t\beta_yy+(\beta_x^2-\beta_y^2)x^2+4\beta_x\beta_yxy-(\beta_x^2-\beta_y^2)y^2\right)\partial_y
\nonumber\\
      &  ~~~-2\gamma_y(t+\beta_xx+\beta_yy)-2\gamma_x(\beta_yx-\beta_xy)\label{righty1y}
\end{align}
\end{subequations}
They satisfy the following non-vanishing commutation relations, with $n,m\in\{0, \pm 1\}$
\BEA
     && [X_n, X_{m}]=(n-m)X_{n+m},                                                     \nonumber\\
     && [X_n, Y_m^x]=(n-m)Y_{n+m}^x \;\; , \;\;  [X_n, Y_m^y]=(n-m)Y_{n+m}^y,          \nonumber\\
     && [Y_n^x, Y_{m}^y]= ~~ [Y_n^y, Y_{m}^x]=(n-m)(\beta_y Y_{n+m}^x+\beta_x Y_{n+m}^y),  \label{mconfcom} \\
     && [Y_n^x, Y_{m}^x]=-[Y_n^y, Y_{m}^y]=(n-m)(\beta_x Y_{n+m}^x-\beta_y Y_{n+m}^y), \nonumber %\\
%     && [Y_m^x, \bar{R}_{xy}]=Y_m^y \;\; , \;\;  [Y_m^y, \bar{R}_{xy}]=-Y_m^x  .       \nonumber
\EEA
\noindent
{\bf Proposition 9:} {\em The set of generators $\left\langle X_{0,\pm1}, Y_{0,\pm1}^x, Y_{0,\pm1}^y\right\rangle$
defined in (\ref{gensimple}, \ref{genaddi})
closes into the Lie algebra $\mathfrak{meta}(1,2)$ if $\beta_x, \beta_y$, are  fixed constants.}

\BLAU{At first sight, this looks as if one could extend this further by including spatial rotations, with a generator
\BEQ \label{xyrotations}
\bar{R}_{xy} :=x\partial_y - y \partial_x + \gamma_x \partial_{\gamma_y} - \gamma_y \partial_{\gamma_x}+
\beta_x \partial_{\beta_y} - \beta_y \partial_{\beta_x}.
\EEQ
which would add the non-vanishing commutators $[Y_m^x, \bar{R}_{xy}]=Y_m^y$ and $[Y_m^y, \bar{R}_{xy}]=-Y_m^x$ 
to the Lie algebra (\ref{mconfcom}). 
However, doing so the the components of $\vec{\beta}$ would have been considered as
{\em variables}. Then objects such as $\beta_x Y_n^x$ on the right-hand-side in (\ref{mconfcom}) 
can no longer be considered as Lie algebra generators.
Hence, it is necessary to give up spatial rotation-invariance and to {\em fix} the values of the components of $\vec{\beta}$
(as we already did above). 
{}From a physical point of view, the absence of rotation-invariance is natural, since the dynamical equation has
a preferred direction (in this work chosen along the $x$-axis).}

\subsubsection{Two-point function}

A simple application of dynamical symmetries is the computation of covariantly
transforming two-point functions. Non-trivial results can be obtained from so-called
`{\it quasi-primary}' scaling operators $\phi(t,z,\bar{z})$,
which transform co-variantly under the finite-dimensional sub-algebra $\langle A_{\pm 1,0}, Y_{\pm 1,0}^{\pm}\rangle$.
Because of temporal and spatial translation-invariance, we can directly write
\BEQ
F(t,z,\bar{z}) = \langle \phi_1(t,z,\bar{z}) \phi_2(0,0,0) \rangle
\EEQ
where the brackets indicate a thermodynamic average which will
have to be carried out when such two-point functions are to computed in the context of a specific statistical mechanics model.
Extending the generators of \BLAU{$\mathfrak{meta}(1,d)$ constructed above} 
to two-body operators, the covariance is then expressed through the Ward identities
$X_0^{[2]} F = X_1^{[2]} F = Y_0^{\pm,[2]} F = Y_1^{\pm,[2]} F = 0$.
Each scaling operator is characterised by three constants $(\delta,\gamma,\bar{\gamma})$.
Standard calculations (along the well-known lines of ortho- or meta-conformal invariance) then lead to\fnref{ftn:calc}
\BEQ \label{gl:2pointCase1}
F(t,z,\bar{z}) = F_0\,\delta_{\delta_1,\delta_2} \delta_{\gamma_1,\gamma_2} \delta_{\bar{\gamma}_1,\bar{\gamma}_2}\:
t^{-2\delta_1} \left(1+\beta \frac{z}{t}\right)^{-2\gamma_1/\beta} \left(1+\beta \frac{\bar{z}}{t}\right)^{-2\bar{\gamma}_1/\beta}
\EEQ
where $F_0$ is a normalisation constant. This shows a cross-over between an ortho-conformal two-point function
when $t\ll z,\bar{z}$ and a novel scaling form in the opposite case $t\gg z,\bar{z}$.
We illustrate this for scalar quasi-primary scaling operators, where $\gamma_1=\bar{\gamma}_1$
\BEQ
F(t,z,\bar{z}) \sim \left\{
\begin{array}{ll} t^{-2\delta_1} \left( \frac{z}{t}\frac{\bar{z}}{t} \right)^{-2\gamma_1/\beta} & \mbox{\rm ~~;~ if $t\ll z,\bar{z}$} 
\\[0.12truecm]
                  t^{-2{\delta}_1} \exp\left[-2\gamma_1 \frac{z +\bar{z}}{t}\right]  & \mbox{\rm ~~;~ if $t\gg z,\bar{z}$}
\end{array} \right.
\EEQ
If the time-difference is small compared to the spatial distance,
the form of the correlator reduces to the one of standard, ortho-conformal invariance.
For increasing time-differences $t$, the behaviour becomes increasingly close to the known one of effectively $1D$
meta-conformal invariance.\footnote{We did not yet carry out explicitly the full algebraic procedure which should in the
$t\gg z,\bar{z}$ limit produce the non-diverging behaviour
$F \sim t^{-2{\delta}_1} \exp\left[-2\beta\gamma_1 \left|\frac{z +\bar{z}}{t}\right|\right]$, see \cite{Henkel16}.}

%%===========================================================================================================================%%
\begin{figure}[hp]
\begin{center}
\includegraphics[width=.65\hsize]{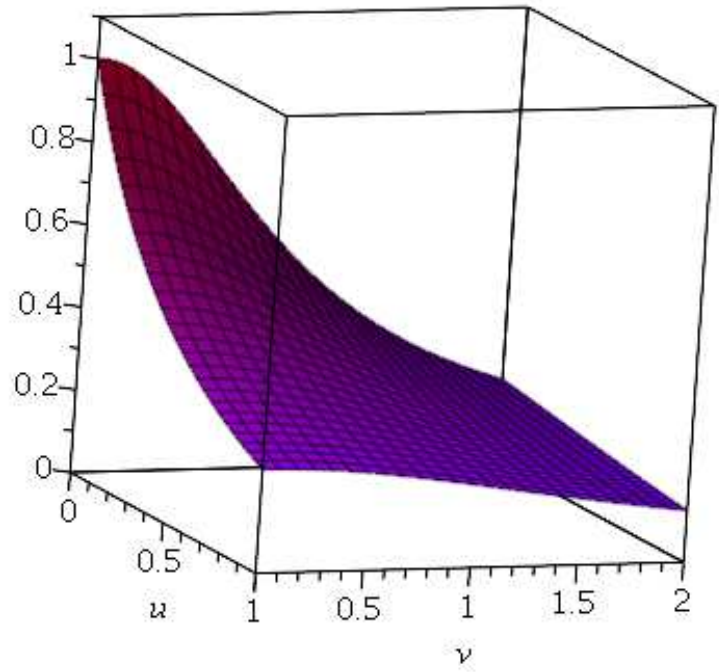} ~~~ \includegraphics[width=.65\hsize]
{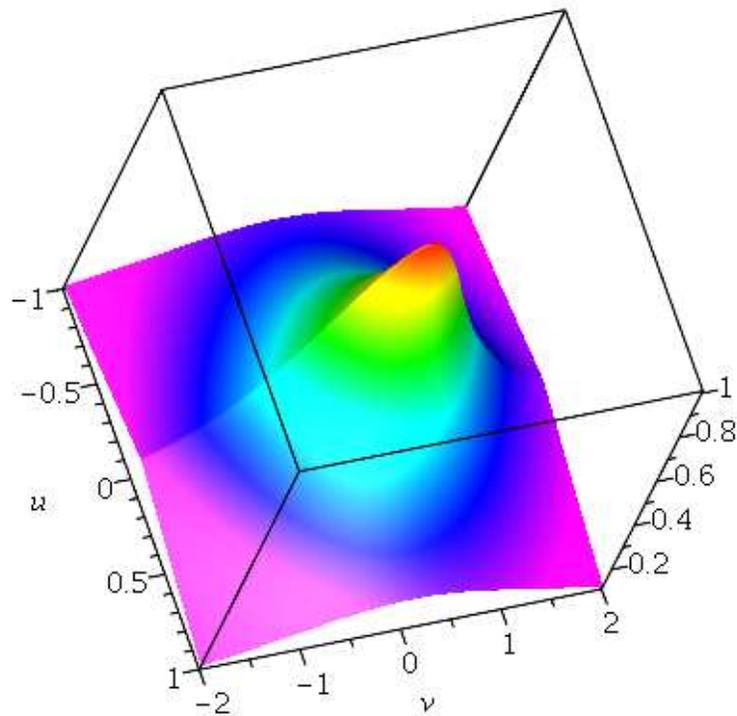}
\end{center}
\caption[fig3]{\small Co-variant meta-conformal two-point function $f$ (\ref{gl:2pointCase1_ext}).
The upper panel shows the quadrant $u\geq 0$, $v\geq 0$,
the lower panel shows the scaling function in the entire plane $(u,v)\in\mathbb{R}^2$.
\label{fig3}}
\end{figure}
%%===========================================================================================================================%%
The two-point function (\ref{gl:2pointCase1}) can be written in the scaling form 
$F(t,z,\bar{z})=t^{-2\delta_1} f(z/t, \bar{z}/t)$.
Using the algebraic construction described in \cite{Henkel15,Henkel15a,Henkel16},
and restricting to the `scalar' case $\gamma_1=\overline{\gamma}_1$ for
notational simplicity, the scaling function $f(u,v)$ can be extended from the
sector $u\geq 0$, $v\geq 0$ to the full plane $(u,v)\in\mathbb{R}$ in the following form (setting $\beta=1$)
\BEQ \label{gl:2pointCase1_ext}
f(u,v) = \left( \left( 1 + |u| \right)^2 + v^2 \right)^{-2\gamma_1}
\EEQ
Figure~\ref{fig3} displays $f(u,v)$. The change from the cusp, characteristic for $1D$ meta-conformal symmetry,
along the $v=0$ axis to the
rounded form of $1D$ otho-conformal symmetry, along  the $u=0$ axis, is clearly seen.

\begin{figure}[tb]
\begin{center}
\includegraphics[width=.7\hsize]{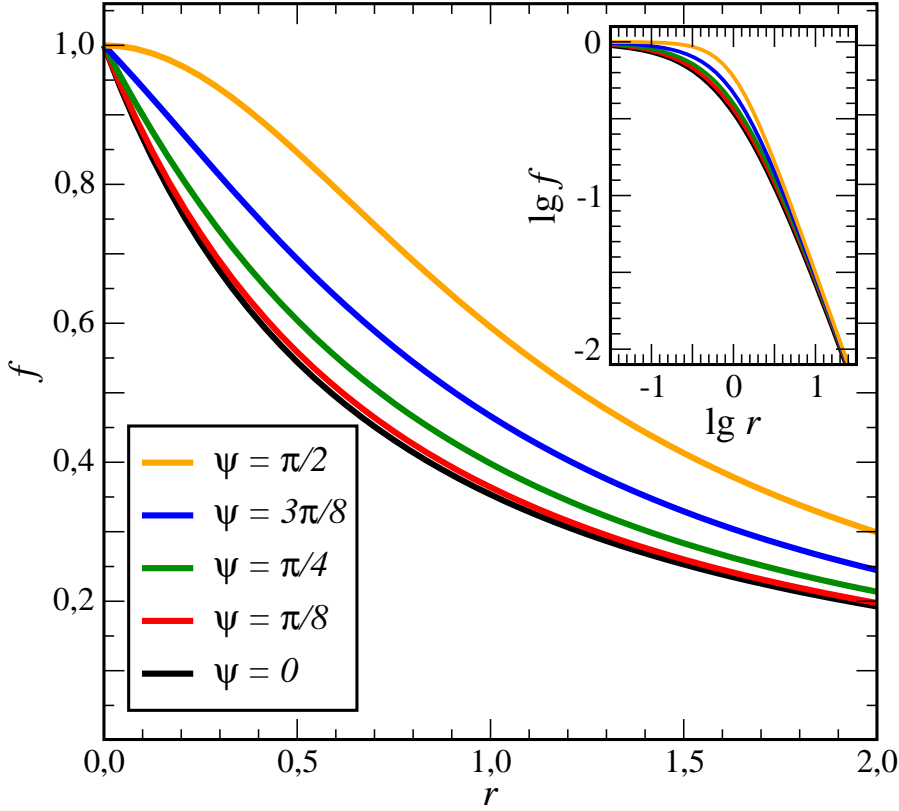}
\end{center}
\caption[fig2]{\small Scaling function (\ref{gl:2pointCase1_ext}) of the co-variant $2D$ meta-conformal two-point function,
in the polar form $f=f(r\cos \psi,r\sin \psi)$ for the angles $\psi=[0,\pi/8,\pi/4, 3\pi/8,\pi/2]$ from bottom to top and with
$\gamma_1=\overline{\gamma}_1=\frac{1}{4}$. The inset displays the same function on a doubly logarithmic scale, 
with $\lg 10^n =n$.}
\label{fig2}
\end{figure}

In figure~\ref{fig2}, the variation of the scaling function (\ref{gl:2pointCase1_ext}) is shown in polar coordinates,
viz. $f=f(r\cos \psi,r\sin \psi)$, over against
the length amplitude $r$,  for fixed values of the angle $\psi$.
The value $\psi=0$ corresponds to the $1D$ meta-conformal case with its characteristic cusp at $r=0$.
The value $\psi=\pi/2$ corresponds to the $1D$ ortho-conformal case with is rounded profile near to $r=0$.
For larger values of $r$, the decay of the scaling function becomes independent of $\psi$. 
The other three quadrants look analogously.

%%===========================================================================================================================%%
%%===========================================================================================================================%%

%
\subsection{The case $p=1/3$}
\BLAU{Again, we refer to tables~\ref{tab3} and~\ref{tab4} for the rendering of the canonical coordinates $(\tau,w,\bar{w})$ 
in terms of the  usual time-space coordinates $(t,x,y)$ and the meta-conformal weights. 
The correspondence between the generators is now as follows
\BEQ \label{glcas2corr}
A_n = \frac{1}{3} X_n + \frac{3}{\beta} Y_n^x \;\; , \;\;
B_n^{\pm} = X_n - \frac{3}{2\beta} Y_n^x \mp \II \frac{3}{2\beta} Y_n^y
\EEQ}

Once more, the co-variant two-point function is build from the {\it quasi-primary scaling ope\-ra\-tors} 
$\phi(t,x,y,\gamma,\bar{\gamma})$, 
\BLAU{where we recall $\gamma=\demi(\gamma_x+\II\gamma_y)$ and $\bar{\gamma}=\demi(\gamma_x-\II\gamma_y)$.} 
Taking into the account the covariance of time- and space-translations, we write
\BEQ
F=F(t,x,y,\gamma_1,\gamma_2,\bar{\gamma}_1,\bar{\gamma}_2)
=\left\langle\phi(t_1,x_1,y_1,\gamma_1,\bar{\gamma}_1)\phi(t_2,x_2,y_2,\gamma_2,\bar{\gamma}_2)\right\rangle,
\EEQ
where $t=t_1-t_2, x=x_1-x_2, y=y_1-y_2$. 
In the canonical coordinates, and writing $\tau=\tau_1-\tau_2$, $w=w_1-w_2$ and
$\bar{w}=\bar{w}_1-\bar{w}_2$, the co-variant two-point function reads, up to normalisation
\BEQ \label{endresult}
F = \tau^{-2\vartheta_1}\, w^{-2\Delta_1}\, \bar{w}^{-2\overline{\Delta}_1}
\EEQ
and with the constraints $\vartheta_1=\vartheta_2$, $\Delta_1=\Delta_2$ and $\overline{\Delta}_1=\overline{\Delta}_2$. 
One may re-express this in the original variables, see table~\ref{tab3}, with a qualitative behaviour quite
similar to the case $p=-1$ treated above.

%%%%%%%%%%%%%%%%%%%%%%%%%%%%%%%%%%%%%%%%%%%%%%%%%%%%%%%%%%%%%%%%%%%%%%%%%%%%%%%%%%%%%%%%%%%%%%%%%%%%%%%%%%%%%%%%%%%%%%%%%%%%%%%
\section{\BLAU{Application: the directed Glauber-Ising chain}}
%%%%%%%%%%%%%%%%%%%%%%%%%%%%%%%%%%%%%%%%%%%%%%%%%%%%%%%%%%%%%%%%%%%%%%%%%%%%%%%%%%%%%%%%%%%%%%%%%%%%%%%%%%%%%%%%%%%%%%%%%%%%%%%

\noindent
\BLAU{We now discuss how a meta-conformal dynamical symmetry is realised 
in the relaxational dynamics of the directed Glauber-Ising chain. 
On an infinitely long chain, Ising spins $\sigma_n=\pm 1$ are attached to each site $n$, such that to each configuration
$\{\sigma\}$ of spins the energy ${\cal H}[\sigma] = - \sum_n \sigma_n \sigma_{n+1}$ is associated. 
The dynamics proceeds through flips of individual spins and is described by a markovian master equation \cite{Tome14}. 
The rates for a flip of the spin $\sigma_n$ is given by \cite{Godreche11,Godreche15a}
\BEQ \label{gl:7.taux}
w_n(\sigma_n) = \demi \left[ 1 - \frac{\gamma}{2}(1-v)\sigma_{n-1}\sigma_n - \frac{\gamma}{2}(1+v) \sigma_n\sigma_{n+1}\right]
\EEQ
%%===========================================================================================================================%%
\begin{figure}[tb]
\begin{center}
\includegraphics[width=.475\hsize]{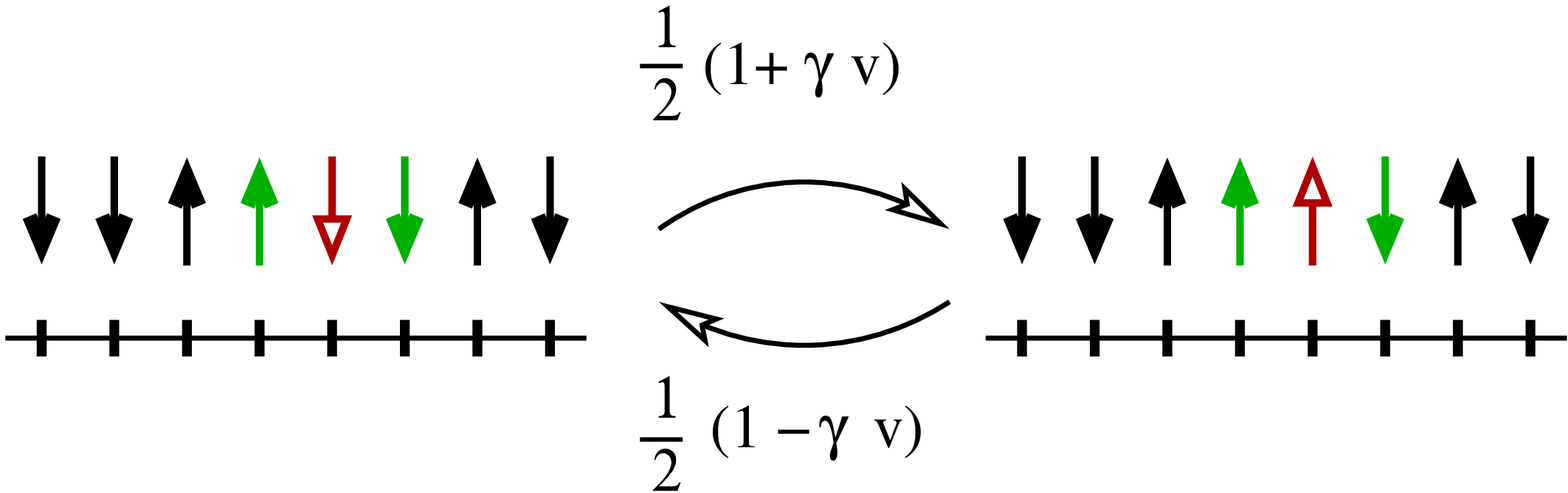} ~~~
\includegraphics[width=.475\hsize]{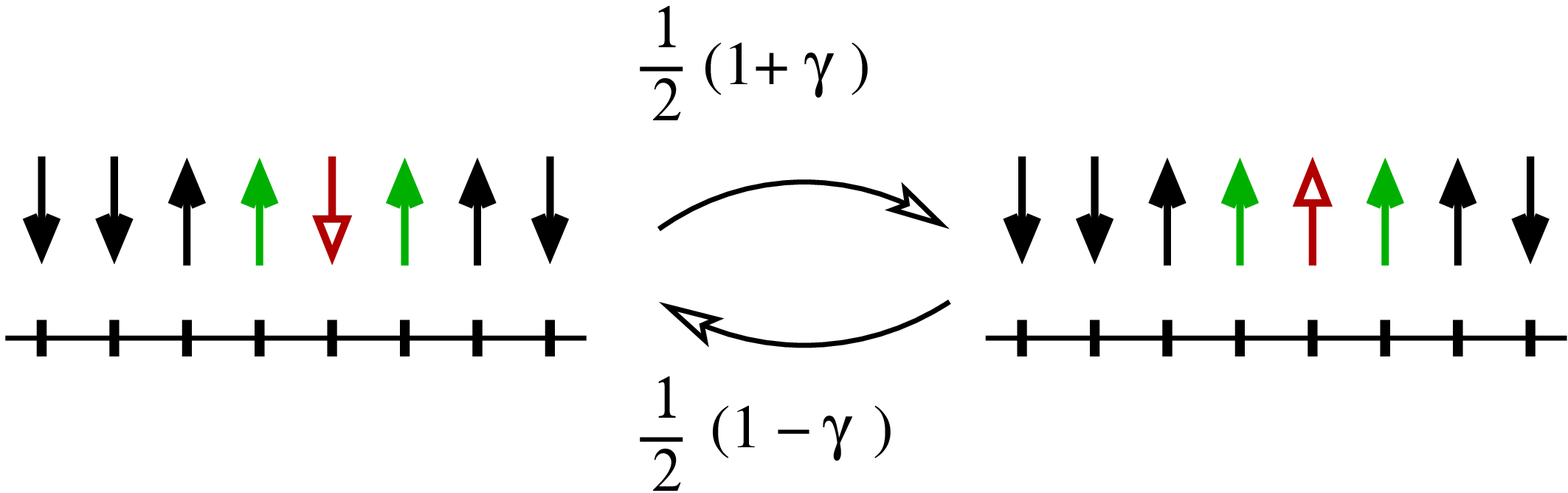}
\end{center}
\caption{\small Dependence of the transition rates on the bias $v$ in the directed Glauber-Ising model. 
The transition rates for
flipping the central spin (open arrow, in \ROT{red}) depend on the orientation of its two nearest neighbours (in \VERT{green}). 
The bias $v$ modifies the rates if (a) the two neighbours have different orientation but cancels if 
(b) the orientation of the two neighbours is the same. \label{figIsingbiaise}}
\end{figure}
%%===========================================================================================================================%%
\noindent
where $\gamma=\tanh(2/T)$ parametrises the temperature and the left-right bias of the dynamics is described by the 
parameter $v$.\footnote{A bias might arise from the effect of an external electric field acting on charged particles or else
particles moving on an inclined lattice in a gravitational field.} The influence of the parameter $v$ on the transition
rates is illustrated in figure~\ref{figIsingbiaise}. 
Such a directed dynamics does no longer obey the condition of detailed balance, although global balance still holds. 
Therefore, with the rates (\ref{gl:7.taux}), the equilibrium Gibbs-Boltzmann state is still 
a stationary state of the dynamics \cite{Godreche11}. 
For either a fully disordered or else a thermalised initial state, the consequences of a non-vanishing bias $v\ne 0$ 
on the long-time
relaxational properties, especially on the precise way how the equilibrium fluctuation-dissipation theorem is broken, have been
studied in great detail \cite{Godreche11,Godreche15a}. Analogous studies have also been carried out in a
$2D$ directed kinetic Ising model \cite{Godreche14,Godreche15b,Godreche18} 
and the directed $d$-dimensional spherical model \cite{Godreche13}. In particular, a $2D$ directed kinetic Ising model quenched
to $T=0$ from a fully disordered initial state shows strong evidence for a relaxational behaviour with a dynamical exponent
$\mathpzc{z}=1$ \cite{Godreche15b}. 
Important observables of interest are the two-time and single-time spin-spin correlators
\BEQ \label{gl:7:corrs}
C_n(t,s) := \left\langle \sigma_n(t) \sigma_0(s) \right\rangle \;\; , \;\;
C_n(t) := C_n(t,t) = \left\langle \sigma_n(t) \sigma_0(t) \right\rangle
\EEQ
where spatial translation-invariance will be admitted throughout.
At present, we shall merely focus on how a meta-conformal dynamical symmetry is realised in this model. 
As we shall see, it will be essential to consider initial states with spatially long-ranged correlations, 
viz. $C_n(0)\sim |n|^{-\aleph}$ for $|n|\gg 1$.\footnote{For {\em unbiased} dynamics with $v=0$, it is known that long-ranged
initial conditions with $\aleph>0$ do not modify the leading long-time relaxation behaviour of the Glauber-Ising chain 
\cite{Henkel04}.} 
}

\BLAU{
{}From the rates (\ref{gl:7.taux}), the equations of motion of the correlators are readily found \cite{Godreche11}
\begin{subequations} \label{gl:7.C}
\begin{align}
\partial_t C_n(t) &= -2(1-\gamma) C_n(t) + \gamma\Bigl( C_{n-1}(t) + C_{n+1}(t) - 2C_n(t) \Bigr) +\delta_{n,0}\ Z(t) 
\label{gl:7.Ct} \\
\partial_{\tau} C_n(\tau+s,s) &= -(1-\gamma) C_n(\tau+s,s) 
                                 +\frac{\gamma}{2}\Bigl( C_{n-1}(\tau+s,s) + C_{n+1}(\tau+s,s) - 2C_n(\tau+s,s) \Bigr) 
                                 \nonumber \\
                              &~~~ +\frac{\gamma v}{2}\Bigl( C_{n+1}(\tau+s,s) - C_{n-1}(\tau+s,s) \Bigr) 
\label{gl:7.Cts}
\end{align}
\end{subequations}
where $\tau=t-s$, the Lagrange multiplier $Z(t)$ is fixed by the condition $C_0(t)=1$, 
one has the compatibility condition $C_n(t,t)=C_n(t)$ and the initial correlator $C_n(0)$ must yet be specified. 
}

\BLAU{
It is known that the requirement of meta-conformal co-variance determines the scaling form of {\em correlators} \cite{Henkel16}, 
rather than response functions as it is the case, e.g. for Schr\"odinger-invariance. 
Concentrating on the correlators (\ref{gl:7:corrs}), from (\ref{gl:7.Ct}) it follows that the single-time correlator $C_n(t)$ 
is independent of the bias $v$ and that one should study the two-time correlators $C_n(t,s)$. 
For illustration, consider first the infinite-temperature limit $\gamma\to 0$ but such that
$\gamma v\to \upsilon$ remains finite \cite{Godreche15a}. Take the continuum limit of (\ref{gl:7.Cts}) and let
$C(\tau+s,s;r)=e^{-\tau}\,\mathscr{C}(\tau+s,s;r)$. This gives the equation
$\bigl(\partial_{\tau} - \upsilon \partial_r\bigr) \mathscr{C}(\tau+s,s;r)=0$, analogous to (\ref{ineq1}), and with the solution
$\mathscr{C}(\tau+s,s;r)=\mathfrak{C}(s;r+\upsilon \tau)$. In the special case $s=0$ of a vanishing waiting time, one has
$C(0,0;r)=\mathscr{C}(0,0;r)=\mathfrak{C}(0;r)$. Hence, for a spatially long-ranged initial correlator $C(0,0;r)\sim |r|^{-\aleph}$,
with $\aleph>0$, the two-time correlator $C(\tau,0;r)\sim e^{-\tau} (r+\upsilon \tau)^{-\aleph}$ 
has indeed the form (\ref{1.12}) predicted by meta-conformal invariance, up to an exponential prefactor. 
}

\BLAU{
We now analyse the long-time behaviour in more detail, and for any temperature $T\geq 0$. 
The equation of motion (\ref{gl:7.Cts}) is solved through a Fourier transformation 
\BEQ
\wit{C}(\tau+s,s;k) = \sum_{n\in\mathbb{Z}} C_n(\tau+s,s)\ e^{-\II n k} \;\; , \;\;
C_n(\tau+s,s) = \frac{1}{2\pi} \int_{-\pi}^{\pi} \!\D k\: e^{\II k n}\: \wit{C}(\tau+s,s;k)
\EEQ
which in Fourier space leads to
\BEQ \label{gl5.7}
\wit{C}(\tau+s,s;k) = \wit{C}(s;k) \exp\left( - \bigl[ 1-\gamma \cos k -\II \gamma v \sin k \bigr] \tau \right)
\EEQ
Tauberian theorems \cite{Fell71} state that the long-time behaviour follows from the form of 
$\wit{C}(\tau+s,s;k)$ around $k\approx 0$. 
Here, we want to look at a `ballistic' scaling regime where $k\tau$ is being kept fixed, 
rather than that regime $k^2\tau = \mbox{\rm cste.}$
typical for diffusive motion. Indeed, for diffusive scaling, the momenta $k\sim \tau^{-1/2} \gg \tau^{-1}$ 
are much larger that the ones to be considered here. From now on, we consider a long-ranged initial correlator of the form 
$C_n(0)\sim |n|^{-\aleph}$, for $|n|\to\infty$ and with $\aleph>0$. A simple explicit form \cite{Henkel98}, which is symmetric in $n$,  
has the required asymptotic behaviour and is normalised to $C_0(0)=1$ reads, along with its Fourier transform \cite{Hansen75} 
\BEQ \label{gl:corrinilong}
C_n(0) = \frac{\Gamma\bigl( |n|+ (1-\aleph)/2\bigr)}{\Gamma\bigl( |n|+ (1+\aleph)/2\bigr)}
\frac{\Gamma\bigl( (1+\aleph)/2\bigr)}{\Gamma\bigl( (1-\aleph)/2\bigr)}
\;\; , \;\; 
\wit{C}(0;k) = \frac{\Gamma\bigl( (1+\aleph)/2\bigr)^2}{\Gamma\bigl( \aleph\bigr)} \left( 2 \sin \frac{|k|}{2}\right)^{\aleph-1}
\EEQ
such that indeed $\wit{C}(0;k)\simeq \wit{C}_0 |k|^{\aleph-1}$, for $|k|$ sufficiently small.}

\BLAU{\underline{\bf (a)} The most simple case arises when the waiting time $s=0$. 
We can directly insert the initial correlator (\ref{gl:corrinilong}) into (\ref{gl5.7}) and read off the two-point correlator 
in the requested scaling limit, and for the range $0<\aleph<1$,
\BEA
C_n(\tau,0) &\simeq& \frac{\wit{C}_0}{2\pi} \int_{\mathbb{R}} \!\D k\: |k|^{\aleph-1} 
\left( 1 - \frac{\gamma}{2} k^2 \tau + \ldots \right)\ e^{\II k (n+\gamma v\, \tau)} \, e^{-(1-\gamma)\tau} \nonumber \\
&\simeq& \frac{\wit{C}_0 \Gamma(\aleph) \cos (\pi\aleph/2)}{\pi} 
         \frac{1}{(n+\gamma v\, \tau)^{\aleph}}\, e^{-(1-\gamma)\tau} \nonumber \\
&=& \frac{\Gamma\bigl((1+\aleph)/2\bigr)^2 \cos (\pi\aleph/2)}{\pi} \frac{1}{(n+\gamma v\, \tau)^{\aleph}}\, e^{-(1-\gamma)\tau}
\label{gl5.8}
\EEA
where the integral is taken from \cite[eq. (2.3.12)]{Gelf64},  see also \cite{Copson65}, and (\ref{gl:corrinilong}) was used. 
The unbiased diffusive terms merely lead to corrections to scaling.  
Eq.~(\ref{gl5.8}) reproduces indeed the prediction (\ref{1.12}) of meta-conformal invariance, 
with $\delta_1=\frac{\gamma_1}{\mu}=\frac{\aleph}{2}$, 
and up to an exponentially decaying prefactor\footnote{Such non-universal exponential factors also arise in other problems, 
for example the number $\mathscr{N}_{\rm saw} \sim e^{N\ln \mathpzc{z}} N^{\bar{\gamma}-1}$ of a self-avoiding random walk ({\sc saw})
of $N\gg 1$ steps contains a non-universal fugacity $\mathpzc{z}$ and an universal exponent $\bar{\gamma}$ \cite{Gennes79}.}
and a choice of scale of spatial distances. Clearly, both the bias $v\ne 0$ as well as
long-ranged initial conditions with $0<\aleph<1$ are necessary ingredients for the meta-conformal dynamical symmetry to arise. 
}

\BLAU{\underline{\bf (b)} For arbitrary waiting times $s>0$, we must now show, under suitable conditions and 
at least for $s$ sufficiently large and for $|k|$ sufficiently small, that 
$\wit{C}(s;k)\simeq \wit{\mathscr{C}}(s)|k|^{\aleph-1}$. 
If that is so, then the two-time correlator $C_n(\tau+s,s)$, see eq.~(\ref{gl5.7}), will
be of the same form as in (\ref{gl5.8}), with a prefactor $\wit{\mathscr{C}}(s)$ which might still depend on the waiting time $s$.}

\BLAU{The proof of this property requires to solve (\ref{gl:7.Ct}). Define the
Laplace transform $\wit{\lap{C}}(p;k) := \int_0^{\infty} \!\D t\: e^{-ps}\: \wit{C}(s;k)$. 
The solution of (\ref{gl:7.Ct}) reads in Laplace-Fourier space
\BEQ \label{gl:Clapfou}
\wit{\lap{C}}(p;k) = \frac{\lap{Z}(p) + \wit{C}(0;k)}{p+2(1 -\gamma \cos k)}
\EEQ
The Lagrange multiplier $\lap{Z}(p)$ is found from the condition $\lap{C}_0(p)= 1/p$. Explicitly
\BEQ \label{gl5.11}
\frac{1}{p} = \frac{1}{2\pi} \int_{-\pi}^{\pi} \!\D k\: \left[ 
\frac{\lap{Z}(p)}{p+2(1-\gamma \cos k)} + \frac{\wit{C}(0;k)}{p+2(1- \gamma \cos k)} \right]
\EEQ
Herein, the first integral can be taken from \cite{Godreche11}. To analyse the second integral, we use again the
explicit form (\ref{gl:corrinilong}) and consider the leading small-$p$ behaviour of
\BEQ \label{glJest}
J(p;\gamma,\aleph) := \frac{\Gamma\bigl((1+\aleph)/2\bigr)^2}{\pi\Gamma\bigl(\aleph\bigr)} 
\int_{0}^{\pi} \!\D k\: \frac{\bigl( 2 \sin k/2\bigr)^{\aleph-1}}{p+2(1 - \gamma \cos k)} 
\stackrel{p\to 0}{\simeq} \left\{ 
\begin{array}{ll} J(0;\gamma,\aleph) & \mbox{\rm ~~;~ if $\gamma<1$} \\
J_{\infty}\, p^{\aleph/2-1} & \mbox{\rm ~~;~ if $\gamma=1$} 
\end{array} \right.
\EEQ
where $J_{\infty} = \left. \Gamma\bigl((1+\aleph)/2\bigr)\Gamma\bigl(1-\aleph/2\bigr)\right/ 2^{\aleph} \sqrt{\pi\,}$. 
For $\gamma<1$, $J(0;\gamma,\aleph)$ is a finite constant. From the constraint (\ref{gl5.11}), and since $\aleph>0$, 
this implies for the leading small-$p$ behaviour of the Lagrange multiplier
\BEA
\lap{Z}(p) &=& \bigl( p+2(1-\gamma)\bigr)^{1/2}\bigl( p+2(1+\gamma)\bigr)^{1/2}\left( \frac{1}{p} - J(p;\gamma,\aleph)\right) 
\nonumber \\
&\stackrel{p\to 0}{\simeq}& \left\{ 
\begin{array}{ll} 2\sqrt{1-\gamma^2\,}\, p^{-1} \bigl( 1 + {\rm o}(p)\bigr) & \mbox{\rm ~~;~ if $\gamma<1$} \\[0.12truecm]
2\, p^{-1/2} \hspace{1.32truecm}\bigl( 1 + {\rm o}(p)\bigr) & \mbox{\rm ~~;~ if $\gamma=1$} 
\end{array} \right.
\label{gl:Zlapasy}
\EEA
where the estimates (\ref{glJest}) for $J(p;\gamma,\aleph)$ were used. We see that the leading behaviour of $\lap{Z}(p)$ 
is independent of the initial condition.}

\BLAU{Using (\ref{gl:Zlapasy}), we now examine the correlator (\ref{gl:Clapfou}) in the asymptotic double limit $p\to 0$ and $k\to 0$. 
Because of the dynamical exponent $\mathpzc{z}=1$ of meta-conformal invariance, 
we expect that this limit should be taken such that $p/k$ is being kept fixed. First, for $\gamma<1$, we find
\BEQ
\wit{\lap{C}}(p;k) \simeq \frac{2 \sqrt{1-\gamma^2\,}\, p^{-1} + \wit{C}_0 |k|^{\aleph-1}}{2(1-\gamma) + {\rm O}(p, k^2)} 
\simeq \sqrt{\frac{1+\gamma}{1-\gamma}\,}\, p^{-1} \bigl( 1 + {\rm o}(1)\bigr) \;\; ;\;\; \mbox{\rm if $\gamma<1$} 
\EEQ
because for $\aleph>0$, the second term in the numerator is less singular than the first one. 
Hence, going back to sufficiently long waiting times $s\gg 1$, we obtain $\wit{C}(s;k) \simeq \sqrt{\frac{1+\gamma}{1-\gamma}\,}$ 
which is constant and
independent of the long-range initial conditions. Hence for $\gamma<1$ there is no meta-conformal invariance 
of the two-time correlator in the limit of large
waiting times. Second, for $\gamma=1$ we have instead 
\BEQ
\wit{\lap{C}}(p;k) \simeq \frac{ 2\, p^{-1/2} + \wit{C}_0 |k|^{\aleph-1}}{p + k^2} 
\simeq \left\{ \begin{array}{ll} 
\frac{\Gamma\bigl((1+\aleph)/2\bigr)^2}{\Gamma\bigl(\aleph\bigr)}\frac{|k|^{\aleph-1}}{p} 
             & \mbox{\rm ~~;~ if $\aleph < \demi$} \\[0.5truecm]
2\, p^{-3/2} & \mbox{\rm ~~;~ if $\aleph > \demi$} 
\end{array} \right.  \;\; , \;\; \mbox{\rm and if $\gamma=1$} 
\EEQ
Hence, if $\aleph<\demi$, we have the leading long-time behaviour 
$\wit{C}(s;k) \simeq \wit{C}_0 |k|^{\aleph-1}$, with $\wit{C}_0$ given in (\ref{gl:corrinilong}), for the single-time
correlator. We have therefore verified a sufficient condition that the form of the two-time correlator $C_n(\tau+s,s)$ is in 
agreement with the expected form (\ref{1.12}) of meta-conformal invariance. 
On the other hand, if $\aleph>\demi$, no clear evidence
for such an invariance is found. Therefore, for large waiting times $s\to\infty$, meta-conformal invariance of the
two-time correlator can only be established under more restrictive conditions than for $s=0$ 
(or $s$ finite and sufficiently small). 
}

\BLAU{We summarise the results of this section as follows.}

\noindent\BLAU{{\bf Proposition 10:} {\em At zero temperature $T=0$, the two-time spin-spin correlator $C_n(\tau,s)$ 
in the directed Glauber-Ising chain, with long-ranged initial correlators of the form $C_n(0)\sim |n|^{-\aleph}$ with 
$0<\aleph<\demi$, takes for large waiting times $s\gg 1$ and large time differences $\tau=t-s\gg 1$ 
the form (\ref{1.12}), predicted by meta-conformal invariance.}
}

\BLAU{While we gave here an example of $1D$ meta-conformal invariance, we point out that the Lie algebra of $2D$ meta-conformal 
transformations is isomorphic to the dynamical symmetry \cite{Henkel17b} of the spatially non-local stochastic process of $1D$ 
diffusion-limited erosion \cite{Krug81} or the terrace-step-kink model \cite{Spohn99,Karevski17}.}

%%%%%%%%%%%%%%%%%%%%%%%%%%%%%%%%%%%%%%%%%%%%%%%%%%%%%%%%%%%%%%%%%%%%%%%%%%%%%%%%%%%%%%%%%%%%%%%%%%%%%%%%%%%%%%%%%%%%%%%%%%%%%%%
\section{Conclusions}
%%%%%%%%%%%%%%%%%%%%%%%%%%%%%%%%%%%%%%%%%%%%%%%%%%%%%%%%%%%%%%%%%%%%%%%%%%%%%%%%%%%%%%%%%%%%%%%%%%%%%%%%%%%%%%%%%%%%%%%%%%%%%%%

{We have explored the construction of time-space transformations, 
with a dynamical exponent $\mathpzc{z}=1$, which may 
have physical applications as dynamical symmetries. Ortho-conformal transformations have been the well-known standard
example of such transformations, with spectacular applications to conformal field-theory, especially in $2D$ equilibrium phase 
transitions. Our main result is stated in table~\ref{tab1}: there are infinite-dimensional
Lie groups of time-space transformations, both for $d=1$ and $d=2$, which contain the same temporal and spatial translations
as well as dilatations, as the orthoconformal group, yet these transformations are in general {\em not} angle-preserving
and hence cannot be ortho-conformal. The relationship between ortho- and meta-conformal transformation for any $d$ 
is stated in (\ref{gl:confmeta}). 
The $1D$ meta-conformal case illustrates the interest in working with representations of the conformal
group which uses non-orthogonal coordinates. For the $2D$ case, the associated Lie algebra is isomorphic to the direct sum
of three Virasoro algebras, rather than two as one is used to from $2D$ ortho-conformal invariance. 
Tables~\ref{tab2}, \ref{tab3} and~\ref{tab4} show how the generic generators (\ref{gl4.3}) 
are related to the physically motivated time-space transformations.} 

\BLAU{The meta-conformal transformations as constructed here are well-known to 
act as dynamical symmetries of a simple linear equation of ballistic transport. 
A new class of applications has been described here: the long-time, large-distance relaxation of non-equilibrium spin systems
whose dynamics contains a directional bias. If in addition sufficiently long-ranged initial  spatial correlations occur, then
the dynamical scaling regime with $\mathpzc{z}=1$ is described by meta-conformal invariance. We have shown this explicitly
for the two-time spin-spin correlator of the directed Glauber-Ising chain, at vanishing temperature and for a decay exponent
$0<\aleph<\aleph_c=\demi$ of the initial spin-spin correlator $C_n(0)\sim |n|^{-\aleph}$.} 

\BLAU{While this kind of application merely uses the finite-dimensional sub-algebra of meta-conformal invariance, 
the full theory based on the infinite-dimensional symmetry remains to be constructed. 
On the other hand, one still must demonstrate that meta-conformal
symmetries arise in systems which are not described by linear equations of motion.  
Previous experience from the phase-ordering kinetics of non-equilibrium spin systems (where $\mathpzc{z}=2$), 
provides evidence that dynamical Schr\"odinger-invariance applies generically \cite{Henkel10}, for example to
kinetic Ising and Potts models, 
although the Schr\"odinger group was originally constructed as the dynamical symmetry of the free diffusion equation. 
Therefore, by analogy a naturally-looking path for identifying meta-conformally invariant systems appears to be the
study of directed spin systems in $d>1$ spatial dimensions. Our results on the Glauber-Ising chain suggest that meta-conformal
invariance might be found for directed systems quenched to temperatures $T\leq T_c$, 
that is below or onto the critical temperature $T_c$. 
The existence of dynamical scaling with $\mathpzc{z}=1$ in such higher-dimensional
models has already been demonstrated \cite{Godreche15b}.}

\newpage

\noindent
{\bf  Acknowledgements:} We warmly thank the organisers of the 10$^{\rm th}$ International Symposium 
``Quantum Theory and Symmetries''
and of the atelier ``Lie Theory and Applications in Physics XII'' in Varna (June 2017)
for the excellent atmosphere, where the main idea for this work arose.
MH gratefully thanks H. Herrmann and his group ``Rech\-ner\-ge\-st\"utz\-te Physik der Werk\-stof\-fe''
at the Institut f\"ur Bau\-stof\-fe (IfB) at ETH Z\"urich (Switzerland)
for warm hospitality during a sabbatical year 2016/17, where many of the ideas presented here were conceived.
This work was supported by PHC Rila and by Bulgarian National Science Fund Grant KP-06-N28/6.

%%%%%%%%%%%%%%%%%%%%%%%%%%%%%%%%%%%%%%%%%%%%%%%%%%%%%%%%%%%%%%%%%%%%%%%%%%%%%%%%%%%%%%%%%%%%%%%%%%%%%%%%%%%%%%%%%%%%%%%%%%%%%%%
%\newpage

\end{document}